\documentclass[conference,compsoc, letterpaper]{IEEEtran}
  \usepackage[caption=false,font=footnotesize]{subfig}
\usepackage{pgfplots}
\usepackage{adjustbox}
\usepackage{listings}
\usepackage{graphicx}
\usepackage{enumitem}
\usepackage{cleveref}
\usepackage{booktabs}
\usepackage{omplst}
\usepackage[hyphens]{url}
\usepackage{todonotes}
\usepackage{fancyhdr}

\newcommand{\highlightForReview}[1]{\textcolor{black}{#1}}

\begin{document}
\thispagestyle{empty}

\par\vspace{10mm}
{\bf Marcos Maro\~nas Bravo} (corresponding author)\\
{Barcelona Supercomputing Center (BSC), Carrer de Jordi Girona, 29, 31, 08034 
    Barcelona}\\
{Fax (+34) 93 413 77 21}\\
{Phone (+34) 93 413 72 45}\\
(\texttt{marcos.maronasbravo@bsc.es})

{\bf Sergi Mateo} \\
{Barcelona Supercomputing Center (BSC), Carrer de Jordi Girona, 29, 31, 08034 
    Barcelona}\\
{Fax (+34) 93 413 77 21}\\
{Phone (+34) 93 413 72 45}\\
(\texttt{sergi.mateo@bsc.es})

{\bf Kai Keller} \\
{Barcelona Supercomputing Center (BSC), Carrer de Jordi Girona, 29, 31, 08034 
    Barcelona}\\
{Fax (+34) 93 413 77 21}\\
{Phone (+34) 93 401 08 40}\\
(\texttt{kai.keller@bsc.es})

{\bf Leonardo Bautista-Gomez} \\
{Barcelona Supercomputing Center (BSC), Carrer de Jordi Girona, 29, 31, 08034 
    Barcelona}\\
{Fax (+34) 93 413 77 21}\\
{Phone (+34) 93 401 63 03}\\
(\texttt{leonardo.bautista@bsc.es})

{\bf Eduard Ayguad\'{e}} \\
{Barcelona Supercomputing Center (BSC), Carrer de Jordi Girona, 29, 31, 08034 
    Barcelona}\\
{Universitat Polit$\grave{e}$cnica de Catalunya (UPC), Carrer de Jordi Girona, 
    1-3, 08034 Barcelona}\\
{Fax (+34) 93 413 77 21}\\
{Phone (+34) 93 413 40 37}\\
(\texttt{eduard.ayguade@bsc.es})

{\bf Vicen\c{c} Beltran} \\
{Barcelona Supercomputing Center (BSC), Carrer de Jordi Girona, 29, 31, 08034 
    Barcelona}\\
{Fax (+34) 93 413 77 21}\\
{Phone (+34) 93 413 77 33}\\
(\texttt{vicenc.beltran@bsc.es})

%
\title{Extending the OpenCHK Model with Advanced Checkpoint Features}

\author{\IEEEauthorblockN{Marcos Maro\~nas\IEEEauthorrefmark{1}, Sergi Mateo\IEEEauthorrefmark{1}, Kai Keller\IEEEauthorrefmark{1}, Leonardo Bautista-Gomez\IEEEauthorrefmark{1}, Eduard Ayguad\'{e}\IEEEauthorrefmark{1}\IEEEauthorrefmark{2} and
        Vicen\c{c} Beltran\IEEEauthorrefmark{1}}
\IEEEauthorblockA{\{marcos.maronasbravo, sergi.mateo, kai.keller, leonardo.bautista, eduard.ayguade, vicenc.beltran\}@bsc.es}
\IEEEauthorblockA{\IEEEauthorrefmark{1} Barcelona Supercomputing Center (BSC)}
\IEEEauthorblockA{\IEEEauthorrefmark{2} Universitat Polit$\grave{e}$cnica de Catalunya (UPC)}
}

\maketitle

This work is an extension of the paper \textit{A Directive-Based Approach to
Perform Persistent Checkpoint\slash Restart} \cite{maronas2017checkpoint}\highlightForReview{,}
published and presented in HPCS2017, Genoa.
\newline
\newline
\newcommand\blfootnote[1]{%
  \begingroup
  \renewcommand\thefootnote{}\footnote{#1}%
  \addtocounter{footnote}{-1}%
  \endgroup
}
\blfootnote{\textcopyright 2020 Elsevier. This manuscript version is made available under the CC-BY-NC-ND 4.0 license \color{blue}{\url{https://creativecommons.org/licenses/by-nc-nd/4.0/}}}
\blfootnote{Published journal article available at \color{blue}{\url{https://doi.org/10.1016/j.future.2020.06.003}}}

%
\begin{abstract}

    One of the
    major challenges \highlightForReview{in using extreme scale systems efficiently is to mitigate the impact of faults}. Application-level
    checkpoint\slash restart (CR) methods provide the best trade-off between
    productivity, robustness, and performance.

    There are many solutions implementing CR at the application level. \highlightForReview{They all}  provide advanced I/O capabilities to minimize the overhead
    introduced by CR. Nevertheless, there is still room for improvement in 
    \highlightForReview{terms of} programmability and flexibility, because end-users must manually serialize
    and deserialize application state using low-level APIs, modify the flow of 
    the application to consider restarts, or rewrite CR code whenever the backend
    library changes.

    In this work, we propose a set of compiler directives and clauses that
    allow users to specify CR operations in a simple way. Our approach
    supports the common CR features provided by all the CR libraries.
    However, it can also be extended to support advanced features \highlightForReview{that are} only
    available in some CR libraries, such as differential checkpointing,
    the use of HDF5 format, and the possibility of \highlightForReview{using}
    fault-tolerance-dedicated threads.

    The result of our evaluation revealed a high increase in
    programmability. On average, we \highlightForReview{reduced} the
    number of lines of code by 71\%, 94\%, and 64\%
    \highlightForReview{for} FTI, SCR, and VeloC, respectively\highlightForReview{, and} no
    additional overhead was perceived using our solution compared to
    using the backend libraries directly. Finally, portability is enhanced \highlightForReview{because}
    our programming model allows \highlightForReview{the} use of any backend \highlightForReview{library} without
    changing \highlightForReview{any} code.

\end{abstract}


\section{Introduction}

Given that exascale systems are expected to be composed of a \highlightForReview{large} number of
components, the mean time between failures (MTBF) will drastically \highlightForReview{decrease} from
the order of days in petascale systems~\cite{reed2004} to the order of hours
in exascale ones~\cite{cappello2014toward}. It is expected that exascale systems
present deep and complex memory/storage hierarchies, hindering the possibility
of non-expert users to exploit \highlightForReview{optimally}. Considering \highlightForReview{these}
facts, the \highlightForReview{high-performance computing} (HPC) research community
is focusing on resilience and fault tolerance \highlightForReview{to} mitigate the impact
of system faults \highlightForReview{more easily. Accordingly,} several
libraries and tools \highlightForReview{are being developed that} leverage low-level details and system nuances to
exploit exascale systems optimally, regardless of \highlightForReview{a} user's expertise. Depending
on the errors \highlightForReview{those systems} address, \highlightForReview{they} can be application-specific~\cite{leo},
algorithm-specific~\cite{du2012algorithm}, or more generic solutions~\cite{blcr}.

One technique that is increasing \highlightForReview{in} popularity is application-level
checkpoint\slash restart \highlightForReview{(CR)}. The main reason is its efficiency in terms of space
and performance compared to other fault-tolerance techniques. \highlightForReview{However}, current
approaches offering application-level \highlightForReview{CR} require considerable effort from the user. Users are in charge of
identifying the application state, serializing and deserializing data for
checkpoints or recovery, and modifying the program flow to check whether the
execution is a restart. Additionally, moving from one system to \highlightForReview{another} may require \highlightForReview{rewriting} the code, at least for tuning.

Among the wide variety of tools and libraries available, three of them stand out
due to their support for multi-level checkpointing: Fault
Tolerance Interface (FTI)~\cite{bautista2011fti}, Scalable Checkpoint and Restart
(SCR)~\cite{scr}, and Very Low Overhead Checkpointing System (VeloC)~\cite{veloc}. These state-of-the-art
libraries also provide optimized I/O capabilities \highlightForReview{and} several redundancy
schemes. Each of them \highlightForReview{provides} its own flexible application programming interface
(API). Nonetheless, compared to other techniques, such as
transparent \highlightForReview{CR} (which requires no participation from the
user but introduces high overhead), they still demand
significant work \highlightForReview{from} the user.

In this paper, we contribute to the aforementioned set of libraries and tools
with an application-level \highlightForReview{CR} mechanism based on compiler
directives. Using compiler directives, we enhance portability and
programmability. Our solution---as FTI, SCR, and VeloC--- supports fail-stop
errors (i.e., process abortions or hardware failures) and soft errors\highlightForReview{, although} undetected errors are not tolerated.

We present the \textbf{OpenCHK} programming model~\cite{openchk} for C/C++ and
Fortran.  Our model is based on compiler directives such as \highlightForReview{in}
OpenMP \cite{openmp}. The sentinel used to recognize the directives and
clauses of the OpenCHK model is \texttt{chk}. \highlightForReview{Currently,} the model supports several
clauses and directives, \highlightForReview{which} are detailed in Section~\ref{sec:new_semantics},
providing users the ability to:

\begin{itemize}
    \item Initialize and finalize the \highlightForReview{CR} environment in \highlightForReview{an easy and
    portable} way.
    \item Easily indicate the data to be protected.
    \item Specify checkpoint conditions (e.g., frequency).
    \item Set checkpoint properties, such as \highlightForReview{identifiers and levels.}
    \item Select among different kinds of checkpoints
          (full/differential).
    \item Avoid the requirement of modifying the natural program flow \highlightForReview{to} check whether the current execution is a restart.
\end{itemize}

In this paper, we extend our initial work~\cite{maronas2017checkpoint} on
pragma-based \highlightForReview{CR} techniques in several directions. First, we
formalize our solution as the OpenCHK programming model. \highlightForReview{Second}, we add new
directives and clauses that \highlightForReview{increase} the expressiveness and \highlightForReview{introduce} new
features to the model, such as fault-tolerance-dedicated threads, differential
checkpoints, and support for the HDF5 format. \highlightForReview{Third, we extend our}
implementation with a new backend library (VeloC) and a useful functionality
called self-iterative data expressions. Last, we also extend the
evaluation to a wider set of production-level scientific applications. More
precisely, we add two new applications\highlightForReview{, xPic and LULESH,} and a new benchmark\highlightForReview{,
Heat 2D simulation}. Additionally, the whole evaluation, including the
benchmarks and applications evaluated in our previous work \highlightForReview{and} the ones
introduced in this work, has been conducted in a new and larger platform,
different \highlightForReview{from} the one used in our previous work. The results of our
evaluation reveal that our solution \highlightForReview{can} maintain efficiency and
robustness, as \highlightForReview{in the} current approaches, while enhancing portability and usability.

The rest of the paper is structured as follows. Section~\ref{sec:background}
contains an introduction to \highlightForReview{the} FTI/SCR/VeloC libraries, as well as the motivations
behind this work. Section~\ref{sec:related} reviews the most relevant related
work. Section~\ref{sec:model} specifies the model semantics and functionalities, and Section~\ref{sec:implementation} provides details of the
implementation. Section~\ref{sec:evaluation} consists of an
evaluation and discussion of our work. Finally, Section~\ref{sec:conclusion}
summarizes the work done and provides some concluding remarks. Future work
proposals are presented in Section~\ref{sec:future}.

\section{Background and Motivations}
\label{sec:background}

\begin{figure}[h]
\centering
\includegraphics[width=\linewidth]{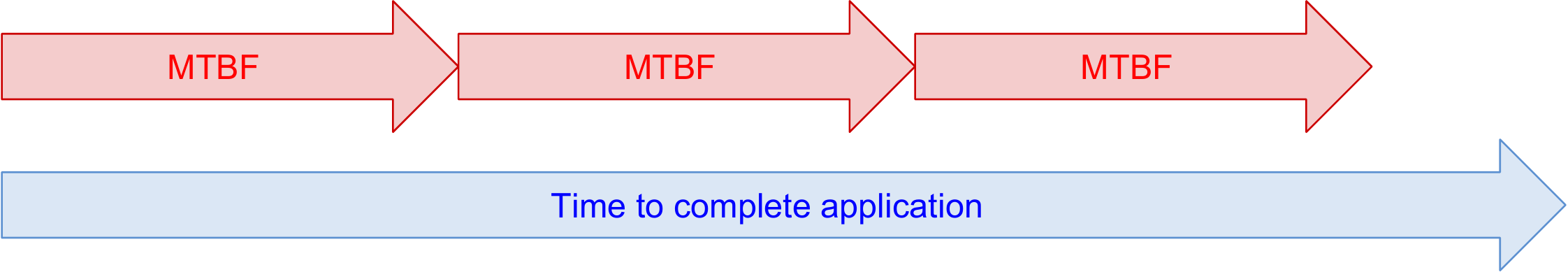}
\caption{\highlightForReview{Long-running applications hardly complete} when MTBF is too small.}
\label{fig:mtbfBTappTime}
\end{figure}

As introduced in the previous section, exascale systems threaten to
jeopardize the successful completion of large HPC
applications. Therefore, fault-tolerance systems become crucial to mitigate the 
impact of errors and guarantee the completion of applications. There \highlightForReview{are} many
HPC applications performing large simulations based on iterative algorithms, 
usually requiring long execution times to \highlightForReview{complete successfully}. Such long
execution times make them more likely to experience system faults. In fact, 
given the increased likelihood of system faults in the exascale era, it is 
possible to find applications requiring more execution time to finish than the 
MTBF of the system. Figure~\ref{fig:mtbfBTappTime} illustrates \highlightForReview{a} scenario where
an application takes more than three times the \highlightForReview{system's} MTBF to complete. Thus,
the application is very unlikely to complete. In this kind of \highlightForReview{scenario},
techniques providing resilience, \highlightForReview{like CR},
become essential.

\highlightForReview{CR} is a widely used technique for saving the full state of
an application in such a way that if an error occurs it can be restored,
allowing the execution to continue from that point instead of from
\highlightForReview{the beginning}. Section~\ref{sec:related} outlines several approaches to provide this
functionality. Our proposal focuses on providing persistent \highlightForReview{CR}. Persistent approaches store the data in a persistent way, usually in
\highlightForReview{a parallel file system (PFS)}. If a node fails and cannot be
recovered, the checkpoints are still accessible.

There \highlightForReview{are} several approaches \highlightForReview{that provide CR}, ranging from
ad-hoc solutions where developers face low-level details \highlightForReview{(such as I/O
operations)}, to libraries offering APIs for abstracting users from such low-level
details and nuances. However, those tools still involve some difficulties,  such
as poor portability and complex APIs.  

\highlightForReview{Regarding} complexity, current approaches require users to (1)
serialize/deserialize data, and (2) modify the natural program flow to check
\highlightForReview{if} a previous checkpoint is available for restart or \highlightForReview{if} the application must
run from scratch. Serialization/deserialization is the \highlightForReview{process} of translating
data structures or objects into a format that can be \highlightForReview{stored efficiently}. For
instance, pointers cannot be stored directly.
Instead, \highlightForReview{their} contents must \highlightForReview{first be} retrieved and then stored. In a restart,
the contents must be read from disk and then \highlightForReview{assigned to} the corresponding pointers.
Regarding the flow, current \highlightForReview{CR} approaches require
users \highlightForReview{to check for checkpoint existence explicitly} and, if any \highlightForReview{exist}, explicitly
ask to recover data. Figure~\ref{fig:complex} shows a very brief example of
these responsibilities. A more complete and real example
can be seen in the appendices, where a full example of a real application is 
included using several state-of-the-art libraries, such as FTI, SCR, and VeloC.
Furthermore, some approaches force users to deal directly with I/O, as in
Appendix~\ref{sec:appendixD}. 

\begin{figure}
\centering
\begin{lstlisting}
    int **data;
    for(int i = 0; i < N; i++){
        data[i] = (int *) malloc(i*sizeof(int));
    }

    // Modifying program flow 
    if(restart_available()) {
        // Deserializing after restart
        int *restarted_data = nullptr;
        size_t restart_size = 0;
        int id = -1;
        for(int i = 0; i < N; i++) {
            restart(&restarted_data, &restart_size, &id);
            assert(id == i);
            assert(restart_size == i);
            memcpy(data[i], restarted_data, restart_size);
        }
    }
    else {
        init_data(data, N);
    }

    // Serializing for checkpoint
    int *cp_data = nullptr;
    size_t cp_size = 0;
    for(int i = 0; i < N; i++) {
        cp_data = data[i];
        cp_size = i;
        checkpoint(cp_data, cp_size, i /* id */);
    }
\end{lstlisting}
\vspace*{-3mm}
\caption{Snippet of pseudocode showing a brief example of the 
         serialization/deserialization process and the modification in
         program flow}
\vspace*{-4mm}
\label{fig:complex}
\end{figure}

\highlightForReview{Second}, the proliferation of several \highlightForReview{CR} libraries with
different interfaces hinders portability between systems. As there is no
standard software stack for \highlightForReview{CR}, it is possible that
different systems offer different \highlightForReview{CR} software. Thus,
writing \highlightForReview{CR} code using native APIs may cause portability
issues when moving to a different system.
In such a situation, \highlightForReview{a} user's options are constrained to \highlightForReview{the following}: (a)
rewriting \highlightForReview{the} code using the interface of the library available in the new
system or (b) installing the original library. Notwithstanding, \highlightForReview{the} installation of
the \highlightForReview{CR} libraries requires deep knowledge of the storage
hierarchy for adequately tuning the installation to maximize
performance. Additionally, it may require special permissions that common users 
do not have. In any case, both options are \highlightForReview{costly} and non-trivial,
affecting portability. 

Using a directive-based approach, the aforementioned problems disappear. The 
model takes several of the user responsibilities (i.e.,
data serialization/deserialization and \highlightForReview{checking} whether a restart is possible) and
\highlightForReview{the users} only \highlightForReview{need} to specify which data must be checkpointed/restored
in a simple way, maximizing programmability. Our solution adds a new
abstraction layer \highlightForReview{with a
unique interface} that enables us to leverage several backend libraries, thereby enhancing portability. The enhancement of portability
comes from enabling developers to use the already tuned installation present in 
every system without changing any code. This approach
enables users to focus on applications, thereby increasing
productivity and portability.

\section{Related work}
\label{sec:related}
We describe different \highlightForReview{CR} approaches
focusing on persistent solutions. We discuss different kinds of checkpointing \highlightForReview{and} examine some checkpointing tools, such as BLCR, FTI, SCR,
and VeloC.

The \highlightForReview{CR} technique consists in regularly storing application
data and restoring it in case of error, thereby benefiting from previous work 
rather than restarting from scratch. For addressing soft errors, data can be 
saved in memory (non-persistent), whereas for hard faults, data must be stored 
persistently in storage.

\highlightForReview{CR} approaches can be organized using several criteria:
application-level or system-level, according to where it is implemented;
persistent or diskless, depending on the \highlightForReview{method} of storing data; and coordinated
or non-coordinated, \highlightForReview{according} to whether process coordination is required
to \highlightForReview{create the checkpoints.}

In coordinated checkpointing, the processes must coordinate to take their 
checkpoints building a global state. In other words, all the processes must \highlightForReview{create}
checkpoints \highlightForReview{simultaneously}. \highlightForReview{This simplifies the recovery process}
because there is no problem with possible rollback propagations. Additionally, 
coordinated mechanisms only need one checkpoint for a successful recovery, 
reducing storage overhead. Non-coordinated checkpointing, in contrast, allows 
processes to \highlightForReview{create} checkpoints at any moment. This is a great advantage because
checkpoints can be \highlightForReview{created} when it is most convenient\highlightForReview{, but,} on a
restart, a \highlightForReview{globally} consistent state must be built \highlightForReview{by }searching the whole set of
saved checkpoints. Therefore, non-coordinated checkpointing may be affected by 
rollback propagation, ending up resuming from the beginning of the execution. 
Thus, overhead grows both in terms of performance and especially storage
space, \highlightForReview{because} each process must keep several checkpoints. CoCheck
\cite{stellner1996cocheck} is an example of coordinated checkpointing, while 
\cite{bronevetsky2003automated} is non-coordinated.

With the objective of removing the main source of overhead, diskless 
checkpointing~\cite{plank1998diskless}, \cite{zheng2004ftc} eliminated 
stable storage from checkpointing. However, non-persistent approaches are less 
resilient than their persistent counterparts, and they \highlightForReview{cannot} tolerate
complete system failures such as power outages. Furthermore, \highlightForReview{they increase}
memory, processor, and network overhead.

\highlightForReview{There are several} persistent checkpointing solutions, providing either
system-level or application-level checkpointing. The strongest point of 
system-level approaches, such as~\cite{duell2002requirements}, 
\cite{roman2002survey}, \cite{duell2005design}, \cite{sankaran2005lam}, or
\cite{blcr} is the transparency: no changes in the application code are required. 
However, \highlightForReview{this comes at the cost of} higher overhead in performance and disk
space compared to application-level solutions.

There are some solutions that are halfway between system-level and 
application-level, such as \highlightForReview{that} proposed by Bronevetsky et al.
\cite{bronevetsky2004application} for shared memory environments. The authors 
present it as an application-level approach, but the user cannot decide which 
data must be checkpointed nor the frequency of the checkpoints. In fact, the 
user can only place some calls to a given method indicating that a checkpoint 
must be taken. Then, the system saves the heap, call stack, and local
and global variables. Given the low degree of freedom provided to the user, it 
cannot be considered a pure application-level solution. Apart from that, most 
applications do not need to save all the data stored by this approach \highlightForReview{for a successful restart, but only a
subset of it}. Thus, overhead is
increased both in performance and disk space usage.

\highlightForReview{There are also a variety of solutions} at application-level~\cite{young1974first},
\cite{duda1983effects}, \cite{plank2001processor}, \cite{daly2006higher}. Some provide single level checkpointing while a few provide multi-level
checkpointing. Applications that store all their checkpoints in the PFS may
introduce \highlightForReview{a large amount} of overhead~\cite{roadrunnertechreport2007},
\cite{schroeder2007}, \cite{oldfield2007modeling}, \cite{reed2004}, 
\cite{schroeder2010large}. Given the gap between the CPU and I/O performance, 
multi-level checkpointing~\cite{gelenbe1976model}, \cite{vaidya1994case} becomes 
essential for reducing overhead. The key is using \highlightForReview{different---and faster--- components}
than PFS, such as RAM disks, local node storage, or SSDs to write the
checkpoints, and moving those checkpoints only when necessary, asynchronously 
and transparently. FTI~\cite{bautista2011fti},
SCR~\cite{scr} and VeloC~\cite{veloc} are multi-level \highlightForReview{CR} solutions.

Those libraries overlap in their multi-level character and their multiple 
redundancy schemes, like partner checkpoints and erasure codes. However, they
differ in the way these schemes are applied. In FTI and VeloC, the cluster 
topology is detected automatically and the appropriate partner nodes for the 
redundancy schemes are selected by the library. In contrast, SCR allows a 
slightly more flexible setup. Besides the standard groups NODE and WORLD, users 
or system administrators may define additional groups (e.g., all nodes that 
share a common power supply). This can be used to increase the likelihood of 
successful recoveries from the various redundancy levels.

VeloC was started as a project to combine FTI and SCR into
a single framework. On the one hand, \highlightForReview{it offers} a \textit{memory-based mode} 
\highlightForReview{that is} very similar to FTI. On the other hand, there is a 
\textit{file-based mode} that behaves much \highlightForReview{like} SCR. However, VeloC is still
missing some features that FTI or SCR support, e.g., different checkpointing 
\highlightForReview{types} (i.e., full checkpoint, differential checkpoint, etc.).

The libraries come with a distinct set of features. For instance, FTI supports 
differential checkpointing\highlightForReview{, and SCR can 
interact} with the running execution and halt the execution \highlightForReview{either} immediately,  
after a certain checkpoint, or at a specific time. The different set of features 
and the assumption that different clusters will provide different 
\highlightForReview{CR} libraries suggest a common interface for these 
libraries which can be used inside HPC applications to run on different systems 
without changing the code. Such an interface is proposed in this paper.

\section{OpenCHK Model}
\label{sec:model}

In this section, we detail the specification of the OpenCHK programming model,
including the directives and clauses supported \highlightForReview{and} the functionalities
provided. The aim of our model is to provide a standard way of coding
\highlightForReview{CR}. We offer a new level of abstraction that hides
implementation details from users, enabling them to focus on the application.
The approach presented in this paper is similar to the one used in some
programming models, such as OpenMP, to exploit parallelism in shared-memory
environments. The rest of this section is structured as follows: \highlightForReview{first}, we
present the directives and clauses of the OpenCHK model, and then we detail the
functionalities offered by the model.

\subsection{Directives and Clauses}
\label{sec:new_semantics}

The model supports four directives. Some of these may also be annotated with
clauses that can modify their \highlightForReview{semantics} in some way. Details on both directives
and clauses are provided as follows.

\begin{enumerate}[label={}]
    \item []\textbf{Directives}

    \item []\textbf{\texttt{init [clauses]}}: The init directive defines the
        initialization of a checkpoint context. A checkpoint context is
        necessary to use the other directives. It accepts the clause:
        \begin{itemize}
            \item \textbf{\texttt{comm(comm-expr)}}: comm-expr becomes the MPI
                communicator that should be used by the user in the checkpoint
                context that is being created. This clause is mandatory.
        \end{itemize}

    \item []\textbf{\texttt{load(data-expr-list) [clauses]}}: This directive
        triggers a load of the data expressed inside the parentheses. The load
        directive accepts the clause:

        \begin{itemize}
            \item \textbf{\texttt{if(bool-expr)}}: The if clause is used as a
                switch-off mechanism: the load will be ignored if the bool-expr
                evaluates to false.
        \end{itemize}

    \item []\textbf{\texttt{store(data-expr-list) [clauses]}}: The store
        directive may request the library to save the specified data. It
        accepts the clauses:

    \begin{itemize}

        \item \textbf{\texttt{if(bool-expr)}}: The if
            clause is used as a switch-off mechanism: the store will be ignored
            if the bool-expr evaluates to false. This clause is useful for
            specifying the desired checkpoint frequency.

        \item \textbf{\texttt{id(integer-expr)}}: Assigns an identifier to the
            checkpoint. This clause is mandatory for the store directive. The
            id of a checkpoint is helpful for later identification of an
            application's progress upon failure. For instance, if users are
            checkpointing steps in a loop, and the id is the loop induction
            variable, users can easily infer which step was last checkpointed.
            However, only the last successful checkpoint is kept.

        \item \textbf{\texttt{level(integer-expr)}}: Selects the checkpoint level
            which is associated with where the data is stored (e.g., local node
            storage, \highlightForReview{PFS}, etc.) and the redundancy schemes
            applied. This clause is mandatory for the store directive. Users
            must \highlightForReview{consider} that the backend libraries provide a different
            number of levels and different behaviors for equivalent levels.
            This is the only parameter that must be tuned
            depending on the underlying backend library. In a future release, we
	    plan to make this clause optional and rely on the parameters
	    passed in the configuration file.

        \item \textbf{\texttt{kind(kind-expr)}}: Selects the checkpoint kind.
            Currently, two kinds are supported. They are \texttt{CHK\_FULL},
            which performs a full checkpoint; and \texttt{CHK\_DIFF}, which
            performs a differential checkpoint.

    \end{itemize}

    \item []\textbf{\texttt{shutdown}}: Closes a checkpoint context.

\end{enumerate}

Figures~\ref{fig:init_shutdown},~\ref{fig:load}, and~\ref{fig:store} show how the
directives and clauses are used in C/C++ and Fortran. \highlightForReview{Specifically},
figure~\ref{fig:init_shutdown} \highlightForReview{shows} how to initialize and shut down a
checkpoint context; figure~\ref{fig:load} shows how to load several types
of data, ranging from \highlightForReview{simple scalars to 2-dimensional arrays}, including
contiguous and non-contiguous regions; and figure~\ref{fig:store} \highlightForReview{is} the
same as the previous figure but for storing instead of loading. However, as it is a
store, we must assign an identifier and a level, as shown in the figure.

\begin{figure}
\centering
\begin{lstlisting}
  // C/C++ syntax
  #pragma chk init comm(mpi_communicator)
  #pragma chk shutdown

  // Fortran syntax
  !$chk init comm(mpi_communicator)
  !$chk shutdown
\end{lstlisting}
\vspace*{-3mm}
\caption{Snippet of code using init and shutdown directives in C/C++ and Fortran}
\vspace*{-4mm}
\label{fig:init_shutdown}
\end{figure}

\begin{figure}
\centering
\begin{lstlisting}
  // Load a) scalar;
  // b) all array elements from 0 to size-1;
  // c) array2 elements from 2 to 4;
  // d) 2dArray elements from 2 to 4
  //    of all the rows from 0 to n-1

  // C/C++ syntax
  #pragma chk load(scalar, array[0;size], \
          array2[2:4], 2dArray[0;n][2:4]) \
          [if(cond)]

  // Fortran syntax
  !$chk load(scalar, array(0:size-1), &
  &        array2(2:4), 2dArray(0:n-1)(2:4) &
  &        [if(cond)]
\end{lstlisting}
\vspace*{-3mm}
\caption{Snippet of code using load directive in C/C++ and Fortran}
\vspace*{-4mm}
\label{fig:load}
\end{figure}

\begin{figure}
\centering
\begin{lstlisting}
  // Store a) scalar;
  // b) all array elements from 0 to size-1;
  // c) array2 elements from 2 to 4;
  // d) 2dArray elements from 2 to 4
  //    of all the rows from 0 to n-1

  // C/C++ syntax
  #pragma chk store(scalar, array[0;size], \
          array2[2:4], 2dArray[0;n][2:4]) \
          kind(CHK_FULL/CHK_DIFF) id(0) \
          level(1) [if(cond)]

  // Fortran syntax
  !$chk store(scalar, array(0:size-1), &
  &        array2(2:4), 2dArray(0:n-1)(2:4) &
  &        kind(CHK_FULL/CHK_DIFF) id(0) &
  &        level(1) [if(cond)]
\end{lstlisting}
\vspace*{-3mm}
\caption{Snippet of code using store directive in C/C++ and Fortran}
\vspace*{-4mm}
\label{fig:store}
\end{figure}

To see a full example, see Appendix~\ref{sec:appendixA}.

\subsection{Functionalities}

\highlightForReview{Our model is intended to standardize} a common interface for the
different \highlightForReview{existing CR} solutions, \highlightForReview{and} we aim to provide the
same functionalities that \highlightForReview{they all} offer but in a generic way.
In what follows, we explain the main functionalities supported in the OpenCHK
model, and how they fit in the \highlightForReview{currently supported backend libraries.}

\subsubsection{Basic Checkpoint\slash Restart}
As a basic functionality, OpenCHK supports checkpoint and recovery of
user-defined application data using the multi-level redundancy schemes of the
backend libraries. Users can define the levels and their respective
checkpoint frequency as desired. Currently, OpenCHK provides a coordinated
\highlightForReview{CR} mechanism because the backend
libraries only support coordinated \highlightForReview{CR}. However, given its
flexibility, OpenCHK could provide non-coordinated \highlightForReview{CR} if
required.

\subsubsection{CP-dedicated Threads}
\label{subsec:cp-dedicated}
This functionality consists of spawning a thread per node that is devoted only
to \highlightForReview{CR} work. By doing so, work related to \highlightForReview{CR} can be \highlightForReview{conducted} in parallel with the application work. This feature may be
useful when running on hybrid CPU-GPU systems or systems using coprocessors
where the main part is executed on GPU/coprocessors and
the CPUs are idle. The idle CPU time can be used to perform \highlightForReview{CR} tasks, relieving the GPU/coprocessors of doing such
tasks and focusing on the actual application work. Overall, resources are better
\highlightForReview{used} in this way, and we can achieve performance gains.
This may increase the memory pressure in some scenarios, but this
effect can be mitigated using local node storage (SSD, NVMe).

\begin{figure}[htbp]
    \centerline{\includegraphics[width=\linewidth]{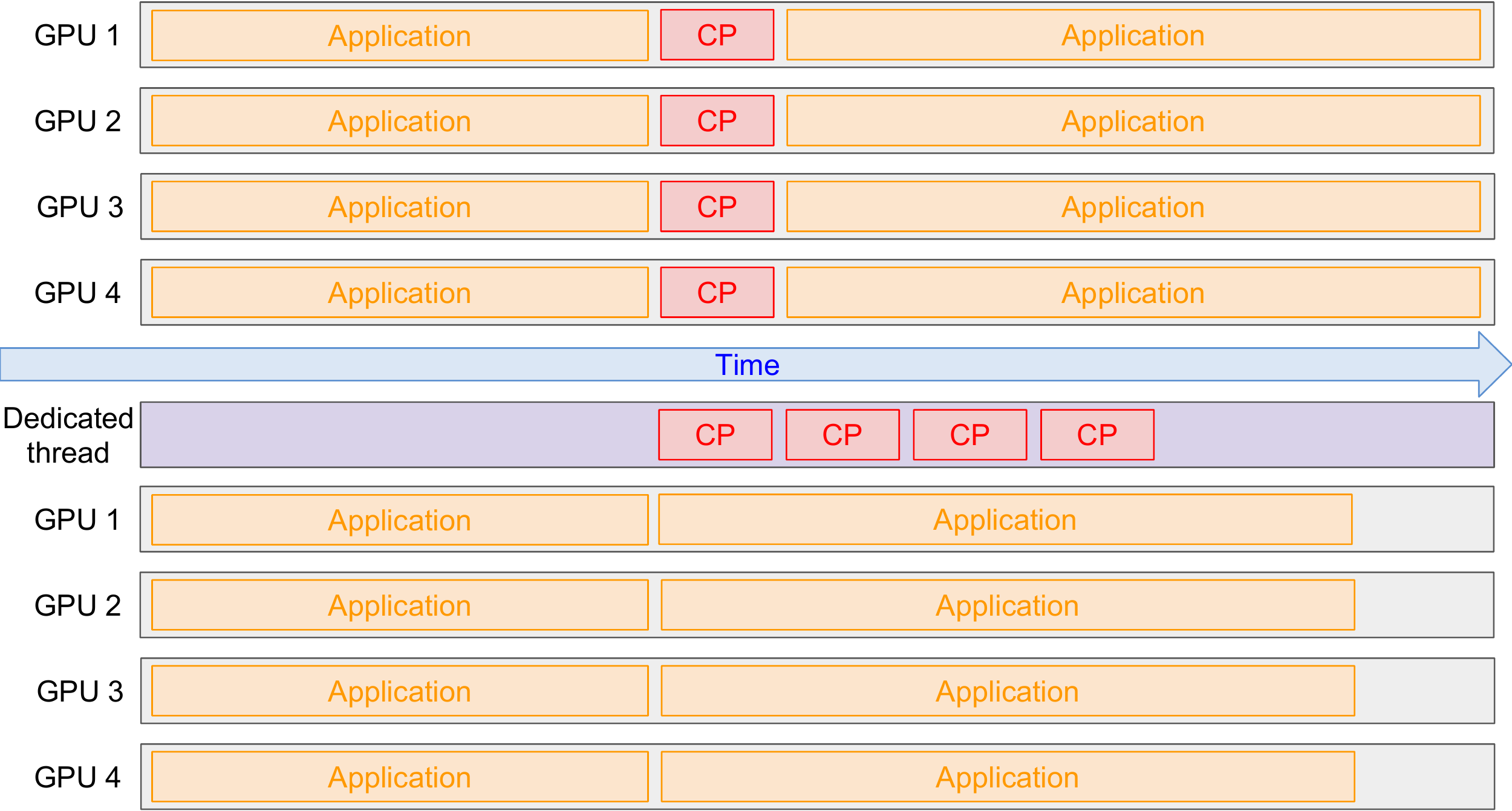}}
    \caption{Comparison between \highlightForReview{a} CP-dedicated threads scheme and a traditional
    scheme.} 
    \label{fig:cp-dedicated-threads}
\end{figure}

Figure~\ref{fig:cp-dedicated-threads} shows a comparison of an application using
the traditional scheme and the same application using this CP-dedicated thread
scheme. A CP-dedicated thread performs all the tasks related to
fault tolerance, while the GPUs can devote their resources to the application.
Up to now, this feature of the model \highlightForReview{is only supported by}
FTI and VeloC.

\subsubsection{Differential Checkpointing}
\label{subsec:diff-cp}
Differential checkpointing is a method that decreases the I/O load of
consecutive checkpoints by updating only those data blocks that have changed
since the last checkpoint was \highlightForReview{created}. Differential checkpointing has also
been \highlightForReview{called} incremental checkpointing~\cite{pbkl:95:lib}.
\highlightForReview{For our purposes}, incremental checkpointing is a different technique
that consists in building a checkpoint \highlightForReview{in} pieces in several separated write
operations that are performed as soon as the data is ready. We plan to support
incremental checkpointing in the future. More information on incremental
checkpointing can be found in Section~\ref{sec:future}. For a more detailed
explanation of our terminology, and why we prefer these terms, please refer
to~\cite{BautistaKellerDcp2019}.

\begin{figure}[htbp]
    \centerline{\includegraphics[width=\linewidth]{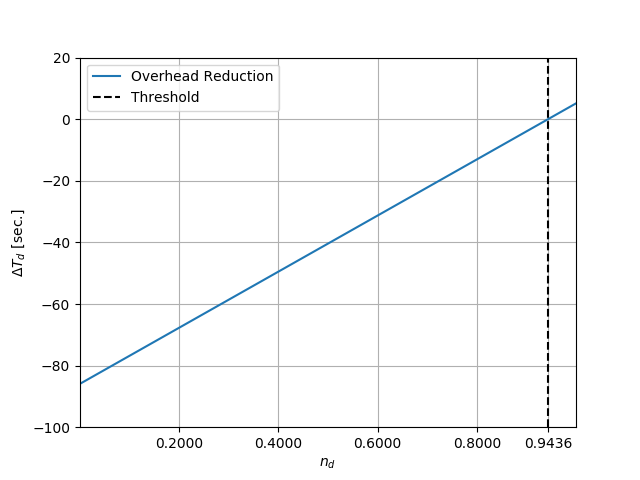}}
    \caption{\highlightForReview{Overhead reduction} with differential
    checkpoint for a certain scenario (2400 processes write
    1 GB per process to the PFS). $n_d$ corresponds to the ratio of dirty data
    blocks to protected data blocks.}
    \label{fig:dcp_overhead}
\end{figure}

Differential checkpointing \highlightForReview{uses} a user-defined block size to evaluate
which sections of the checkpoint have changed. This granularity is important
and offers a trade-off: smaller block sizes can capture with higher precision
small changes in the dataset, which allows the backend to only update small
sections and avoid having to rewrite data that has not changed. However,
performing hash calculations on many small blocks and performing many small
writes to the storage can \highlightForReview{reduce} performance. On the other hand, large
blocks are more suited \highlightForReview{to} file system performance and lead to a reduction in
the number of hashes to be calculated, but they also lead to more unchanged data
having to be rewritten. Besides the block size, the performance of differential
checkpointing also depends on the application itself. Applications that update
entire datasets at every iteration are not well suited for differential
checkpointing. Applications in which only parts of the dataset change,
might get more benefits from differential \highlightForReview{checkpointing}.

For instance, FTI has recently demonstrated ~\cite{BautistaKellerDcp2019}
that applications updating less than 95\% of their protected data due to changes
within two consecutive checkpoints will be able to reduce their
checkpoint overhead using differential checkpointing. In other words, a
reduction in I/O size by as little as 5\% already \highlightForReview{shows} significant benefits
through differential checkpointing. The overhead reduction
depends linearly on the update rate. The slope of the regression characterizes
this linear relationship. The ratio between the I/O rate and the hashing
determines the slope. If the hashing is expensive
and the I/O rate is low, the overhead reduction will be low even for low update
rates, whereas, if the hashing is cheap and the I/O rate high, even high update
rates could obtain a benefit. It was demonstrated for \highlightForReview{the} LULESH and xPic
applications that differential checkpointing can reduce the checkpointing overhead
by 35\% and 62\%, respectively~\cite{BautistaKellerDcp2019}.

Given the model presented in~\cite{BautistaKellerDcp2019}, we can
estimate the performance benefits for a certain scenario.
Figure~\ref{fig:dcp_overhead} shows the behavior of the overhead
depending on the differential data ratio, $n_d$, corresponding to the
ratio of dirty data blocks to protected data blocks. The \highlightForReview{x-axis}
represents the value of the differential data ratio, while the \highlightForReview{y-axis}
represents the increment of time that differential checkpointing
introduces with respect to a common full checkpoint.  Thus, positive
values are additional overhead introduced by differential checkpointing
with respect to a common full checkpoint, while negative values are
benefit against a common full checkpoint. The presented scenario comprises 2400
processes that write 1 GB per process to the PFS. The
time needed to complete a full checkpoint on all ranks is about 88
seconds. The threshold is at around 95\%, which is the
point where differential checkpointing and common full checkpointing
introduce the same amount of overhead. If the differential data ratio is
above this threshold, differential checkpointing introduces a penalty of
up to 5 \highlightForReview{s} compared to a common full checkpoint. In other
words, if more than 95\% of the data must be checkpointed, differential
checkpointing introduces more overhead than a common full checkpoint.
Nevertheless, if the differential data ratio is below the threshold,
there is a benefit. The overhead reduction is about 9 seconds for every 10\%
of data that we do not have to write due to differential checkpointing.
Thus, differential checkpointing becomes very quickly beneficial for
updates below 95\%. \highlightForReview{Of} all the checkpointing libraries studied in this work, the only backend library supporting this functionality is
FTI.

\subsubsection{HDF5 support}
\label{subsec:hdf5}
HDF5\cite{hdf5} allows \highlightForReview{the structuring of} datasets inside of groups and to order
groups hierarchically \highlightForReview{as in} a file system. The dimensionality of the
datasets can also be stored inside the file. HDF5 provides a vast functionality
\highlightForReview{to} archive scientific data inside a file \highlightForReview{in} persistent storage. In
addition to this, HDF5 is optimized for both sequential and parallel I/O.

Our model allows checkpoints to be stored using the HDF5 file format. The
protected datasets that serve for the successful restart are written in this
format so that users can use any tool that is capable of interacting with HDF5
files for scientific analyses. Thanks to this feature, resiliency and
scientific data analysis can be merged into one single I/O operation. Interacting with HDF5 files can be relatively complex and not always
intuitive. Therefore, this feature to support HDF5 files enhances the
flexibility of OpenCHK.

Figure~\ref{fig:hdf5ex} shows an example of the structure of a checkpoint file
in the HDF5 format. The file contains three protected variables: Dataset\_0 and
Dataset\_1, which are scalars, and Dataset\_2, which is an array. Figure
\ref{fig:hdf5-pseudo} provides two different codes able to obtain a checkpoint
like the one in Figure~\ref{fig:hdf5ex}. The first one uses OpenCHK while the
second \highlightForReview{uses} the native HDF5 API. Using OpenCHK, users just
need to use the store directive, indicating the data to be stored along with an
id and a level. Using the native HDF5 API, they \highlightForReview{must} create a dataspace per
variable, indicating the size and shape; create a dataset per variable
indicating the data elements, layout, and some other information necessary to
write, read, and interpret the data; write the data, and, finally, close the
datasets and dataspaces. Looking at the codes, it is possible to see how OpenCHK
reduces the complexity \highlightForReview{compared to} the native implementation.

\begin{figure}
\centering
    \begin{lstlisting}[numbers=none]
GROUP "/" {
   DATASET "Dataset_0" {
      DATATYPE  H5T_STD_I32LE
      DATASPACE  SIMPLE { ( 1 ) / ( 1 ) }
      DATA {
      (0): 1
      }
   }
   DATASET "Dataset_1" {
      DATATYPE  H5T_STD_I32LE
      DATASPACE  SIMPLE { ( 1 ) / ( 1 ) }
      DATA {
      (0): 1
      }
   }
   DATASET "Dataset_2" {
      DATATYPE  H5T_STD_I8LE
      DATASPACE  SIMPLE { ( 22279025 ) / ( 22279025 ) }
      DATA {
      (0): 50, 50, 32, 115, 101, 114, 105, 97, 108, 105,

      ...

      (22279021): 0, 0, 0, 0
      }
   }
}
\end{lstlisting}
\vspace*{-3mm}
    \caption{HDF5 checkpoint file structure using OpenCHK.
    The protected data consists of 2 scalars and one array.}
\vspace*{-4mm}
\label{fig:hdf5ex}
\end{figure}

\begin{figure}[h!]
\centering
\begin{lstlisting}[frame=tb]
// OpenCHK implementation
#pragma chk store(data_ptr[0], data_ptr[1], \
        data_ptr[2][0;N]) id(0) level(1)

// HDF5 Native implementation
hid_t       dataset_id[3], dataspace_id[3];
hsize_t     dims0[1], dims1[0], dims2[0];

dims0[0] = 1; dims1[0] = 1; dims2[0] = N; 
dataspace_id[0] = H5Screate_simple(1, dims0, NULL);
dataspace_id[1] = H5Screate_simple(1, dims1, NULL);
dataspace_id[2] = H5Screate_simple(1, dims2, NULL);

dataset_id[0] = H5Dcreate2(file_id, "/Dataset_0", ... );
dataset_id[1] = H5Dcreate2(file_id, "/Dataset_1", ... );
dataset_id[2] = H5Dcreate2(file_id, "/Dataset_2", ... );

H5Dwrite(dataset_id[0], ... , data_ptr[0]);
H5Dwrite(dataset_id[1], ... , data_ptr[1]);
H5Dwrite(dataset_id[2], ... , data_ptr[2]);

status = H5Dclose(dataset_id[0]);
status = H5Dclose(dataset_id[1]);
status = H5Dclose(dataset_id[2]);
status = H5Sclose(dataspace_id[0]);
status = H5Sclose(dataspace_id[1]);
status = H5Sclose(dataspace_id[2]);
\end{lstlisting}
\vspace*{-3mm}
    \caption{Code snippet to produce an HDF5 file with a structure similar to
    the one shown in figure~\ref{fig:hdf5ex}, on top with OpenCHK and below
    with native HDF5 routines.}
\vspace*{-4mm}
\label{fig:hdf5-pseudo}
\end{figure}

\section{Implementation Details}
\label{sec:implementation}

This section provides some insight into the implementation details of our
proposed solution. We provide our implementation of the model on top of
the Mercurium C/C++ and Fortran source-to-source compiler~\cite{mercurium} and
the Transparent Checkpoint Library (TCL)~\cite{tcl} intermediate library.
Currently, we support FTI, SCR, and VeloC as backend libraries.

Following, we detail the architecture of our implementation, the changes
effected at the Mercurium compiler level, and the implementation of TCL.

\subsection{Architecture}

\begin{figure*}[h]
\centering
\includegraphics[width=.7\linewidth]{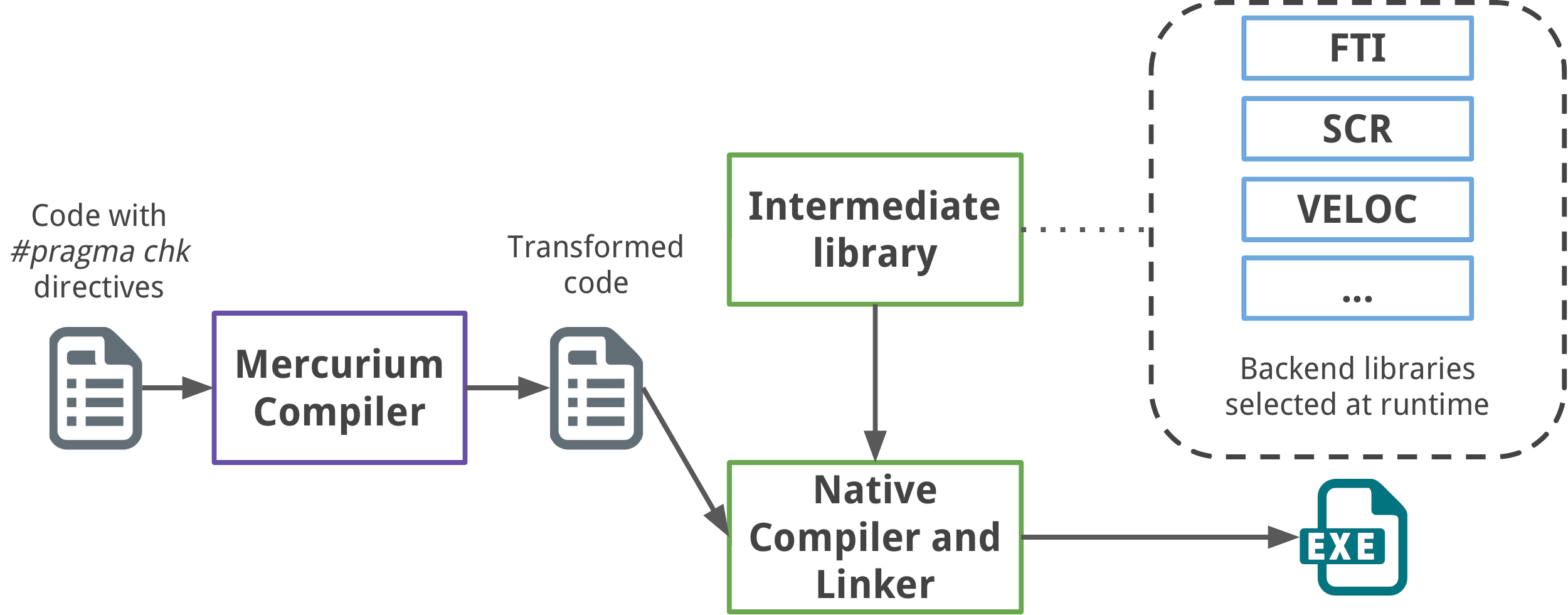}
\caption{Three-layer architecture.} 
\label{fig:3layer}
\end{figure*}

We have designed an implementation based on three components: a compiler
(Mercurium) that translates directives and clauses into calls to an
intermediate library, an intermediate library (TCL) which \highlightForReview{oversees}
forwarding the user-requested actions to the adequate backend library, and
several backend libraries. Figure~\ref{fig:3layer} shows our three-layer
architecture. This approach allows us to extend
the model to support new features \highlightForReview{as} the backend libraries evolve. \newline

\subsection{Mercurium}  
\label{subsec:mcxx}

For supporting the OpenCHK model, Mercurium \highlightForReview{must} process the OpenCHK
directives and clauses to enable the application-level \highlightForReview{CR} functionalities. One of the main duties of Mercurium is processing
these directives and clauses \highlightForReview{to} transform them into calls to TCL. Following,
we detail the compiler transformations done for each of the directives and
clauses.

\begin{itemize}
    \item \textbf{\texttt{chk init [clauses]}}: The compiler triggers the initialization
    of TCL. Clauses accepted:
    \begin{itemize}
        \item \textbf{\texttt{comm(comm-expr)}}: Mercurium passes comm-expr,
            which is a pointer, to TCL, \highlightForReview{which} sets the MPI communicator
            that the user should \highlightForReview{use} in the checkpoint context.
    \end{itemize}

    \item \textbf{\texttt{chk load(data-expr-list) [clauses]}}: The compiler
        introduces a call to TCL when it finds this statement informing \highlightForReview{it} of the
        start of a restart. Then, Mercurium performs several calls registering
        the data to be restarted. \highlightForReview{Accordingly}, the compiler must also send some
        information about the data to be restored, such as the sizes and the
        pointers, for overwriting the current data with the recovered \highlightForReview{data}. The
        compiler extracts all this information from the data specified and its
        own knowledge of the program symbols. When all the data is registered,
        Mercurium calls a TCL method that performs the restart. This is
        part of the deserialization process that \highlightForReview{would otherwise} be done by the user. Additionally, this directive implies a transparent way of
        checking whether a restart exists, \highlightForReview{which would also be done by
        a user modifying the program flow in other approaches.}
        Clauses accepted:
    \begin{itemize}
        \item \textbf{\texttt{if(bool-expr)}}: The calls to TCL are only
        effective when the condition expressed in this clause is true. This
        means that none of the calls are done if the condition is not satisfied.
    \end{itemize}

    \item \textbf{\texttt{chk store(data-expr-list) [clauses]}}:
    The compiler does very similar things \highlightForReview{as} those
    performed for the load directive. The compiler \highlightForReview{oversees} doing a part
    of the data serialization, \highlightForReview{equivalent} to the deserialization process \highlightForReview{conducted} for the load directive. The only difference is that the action being
    performed is a checkpoint instead of a restart, \highlightForReview{meaning} some additional
    information must be passed. In the first call, the one that notifies a
    checkpoint is starting, Mercurium adds the kind, id, and level of the
    checkpoint. This information is extracted from the following clauses:
    \begin{itemize}
        \item \textbf{\texttt{if(bool-expr)}}: The calls to TCL are only
            effective when the condition expressed in this clause is true.
            Thus, no calls are \highlightForReview{executed} if the condition is not satisfied.
    \item \textbf{\texttt{id(integer-expr)}}: The checkpoint that is being
            performed has the identifier set in this clause. It is mandatory to
            specify an identifier.
    \item \textbf{\texttt{level(integer-expr)}}: The checkpoint that is being
            performed will be written at the specified level. It is mandatory to
            specify a level.
    \item \textbf{\texttt{kind(kind-expr)}}: Chooses the kind of checkpoint to
            be done between full or differential, which are the \highlightForReview{currently
            supported options}. The default value is full.
    \end{itemize}

    \item \textbf{\texttt{chk shutdown}}: The compiler triggers the finalization
    of TCL.

\end{itemize}

The order specified in the
load/store clauses \highlightForReview{is critical}. The compiler forwards the data to TCL in the very same order
set by the user when writing the load/store clauses. Thus, if the order of the
data in loads and stores does not match, there may be problems when recovering
data from \highlightForReview{a} restart.

Also, it is important to mention that the three back-end
libraries are capable  of deciding the checkpoint level automatically based
on the configuration file. However, our current implementation in OpenCHK does
require the user to supply the checkpoint level. This is a small limitation that will
be lifted in future versions of OpenCHK.

A further duty of Mercurium, which is crucial for improving code
programmability, is extracting the information that must pass to TCL from the
annotations done by the user. In most cases, backend libraries need the base
address and the size of data to perform \highlightForReview{CR}. For non-array
expressions, this is just the size of the type of the data. However, for more
complex data structures, \highlightForReview{like} array expressions, we may need additional
information, such as the accessed bounds of each dimension, the size of each
dimension, and the base element type of the data structure.

All the aforementioned information required by the backend libraries is only
retrievable by a tool with a full understanding of the supported programming
languages---C, C++, and Fortran--- such as the Mercurium compiler. \highlightForReview{Additionally},
this is an error-prone task. Thus, automatizing \highlightForReview{it}
minimizes the possibility of error while reducing debugging time. Apart from
that, it prevents developers from writing boilerplate code.

A further functionality added by Mercurium is \textit{self-iterative data
expressions}. Figure~\ref{fig:multideps} shows an example of this. This is a
kind of \texttt{for} loop inside the load/store clauses, \highlightForReview{which} allows iterating
over data structures instead of writing the data one by one, simplifying users'
work. Self-iterative data expressions are useful in
scenarios where \highlightForReview{many} elements of a data structure must be
checkpointed/loaded and users must write it \highlightForReview{manually}. In \highlightForReview{these} cases, annotating
the data that must be checkpointed/restored becomes a tedious task. \highlightForReview{It also}
becomes error-prone, because writing \highlightForReview{the same code several times}, changing only a
few characters, may cause errors to be more difficult to find. Self-iterative
data expressions enable users to perform this work in a much easier way.

\begin{figure}
\centering
\begin{lstlisting}
  // Self iterative data expression
  #pragma chk store({data[i], i=0;4})

  // Equivalent
  #pragma chk store(data[0], data[1], data[2], data[3])
\end{lstlisting}
\vspace*{-3mm}
\caption{Snippet of code using self-iterative data expressions}
\vspace*{-4mm}
\label{fig:multideps}
\end{figure}

To adequately point out the importance of the compiler in the OpenCHK model, we
have performed an analysis of the code complexity before and after the compiler
transformation. For that purpose, we used the Lizard~\cite{lizard}
tool to compute the cyclomatic complexity (CC) metric and
SLOCCount~\cite{sloccount} to calculate the development effort estimate (DEE).
Table~\ref{tab:complexity} shows the CC for a 2D heat simulation code
performing \highlightForReview{CR} using OpenCHK before (BT) and after (AT)
compiler transformation. Moreover, it also presents the CC for the same code
when using native FTI, SCR, and VeloC to perform \highlightForReview{CR}. As
can be seen, OpenCHK remains the simplest before the compiler transformation
in both metrics. After the compiler transformation, its CC grows \highlightForReview{by} 1 point while
its DEE more than \highlightForReview{doubles}, becoming the most complex version.

The CC is higher for the native libraries because this metric is affected by the
number of different paths that a program can take. For instance, each additional
\highlightForReview{\texttt{if}} increases the CC. The DEE, in contrast, is affected by the size of the code.
In that case, it is important to highlight that Mercurium generates very verbose
code when transforming code, so the size of it \highlightForReview{quickly becomes large}, affecting
the DEE.

\begin{table}[]
\centering
\caption{Cyclomatic complexity of a 2D heat simulation using OpenCHK (before
    and after compiler transformation), FTI, SCR, and VeloC.}
\label{tab:complexity}
\begin{tabular}{@{}llllll@{}}
\toprule
                   & \textbf{BT} & \textbf{AT} & \textbf{FTI}  & \textbf{SCR}  & \textbf{VeloC} \\ \midrule
\textbf{CC}      & 10.0 & 11.0 & 15.0 & 36.0 & 13.0            \\
\textbf{DEE (Person-Months)}      & 0.34 & 0.76 & 0.36 & 0.51 & 0.35            \\ \bottomrule
\end{tabular}
\vspace*{-4mm}
\end{table}

\subsection{TCL}
\highlightForReview{To} maximize the portability of our approach, TCL \highlightForReview{must} process the
information passed by Mercurium and forward it to the adequate backend library.
This way, we enable users to write code agnostic from the backend library while
allowing them to use any of the supported backend libraries. TCL is responsible
for adequately formatting the information for each backend library and calling
the appropriate methods to perform the user-requested actions,
depending on the backend library chosen by the user.

\highlightForReview{Additionally}, this library, in collaboration with the Mercurium compiler,
serializes and \highlightForReview{deserializes} the data. The serialization and deserialization
process is tedious and error-prone for users. Using our mechanism, users
can perform it with little or no effort.

\highlightForReview{Additionally}, TCL prevents users from modifying the natural program flow to check
whether a restart has to be done. The library does it transparently.
If a checkpoint is available and a restart can be done, it
recovers data. In consequence, codes are cleaner and more readable.

\highlightForReview{The mechanism we propose is easily extensible}. Our first implementation of TCL was providing support for
only FTI and SCR. Now VeloC has been added as a third backend library to TCL, and
the addition of \highlightForReview{others} would be straightforward.

\section{Evaluation}
\label{sec:evaluation}

\highlightForReview{In this section we compare} the performance of our approach
with natively using the backend libraries that we use in our model.
The structure of this section is as follows: \highlightForReview{first}, we describe the
methodology \highlightForReview{used}; then, we detail the environment in which the experiments
were conducted, as well as the benchmarks and applications used. Finally, we
provide the evaluation and \highlightForReview{discuss} the results.

\subsection{Methodology}
\label{subsec:methodology}
\highlightForReview{First, the objectives of our evaluation are}
(1) demonstrating that our approach \highlightForReview{does} not add significant overhead compared to
using any of the supported backend libraries \highlightForReview{natively}, and (2) showing the
improvement \highlightForReview{that our mechanism provides in terms of code productivity.}

For (1), we \highlightForReview{designed} an experiment in which we launch
a first run of an application/benchmark, with an injected fault. Then, we
restart the application/benchmark from the last checkpoint until
successful completion. The whole process, from the first
run until successful completion, including the recovery, is
timed. We take measurements using \highlightForReview{both} OpenCHK with a given backend library and using the same backend library \highlightForReview{natively}. Then, the OpenCHK time is divided by
the native library time. \highlightForReview{To} demonstrate that no significant overhead is
introduced, the resulting number from the quotient should be 1 or close to 1.

Regarding (2), as there is no standard metric for measuring \highlightForReview{programmability}, we
have decided to consider the number of \highlightForReview{source} lines of code (SLOC) required to express
the \highlightForReview{CR} functionality. Thus, we compare the number of code
lines required using native APIs with the number of code lines required with
OpenCHK.

We will use the following nomenclature for our experiments:
\begin{itemize}
    \item \texttt{FTI/SCR/VeloC}. This version is an implementation performing
        application-level \highlightForReview{CR} directly using the APIs
        provided by FTI, SCR or VeloC.

    \item \texttt{OpenCHK}. In this version, the application-level \highlightForReview{CR is conducted} by the mechanism proposed in this work.
\end{itemize}

\highlightForReview{We obtained the results of all our experiments}
by averaging the execution times of 5 different runs per version.

The executions of the evaluation \highlightForReview{were} run with 50 MPI processes, whenever it was
possible. Nevertheless, there are some
applications/benchmarks that \highlightForReview{constrain} the number of processes to be used. For
each of those applications, we specify the number of processes used. We
consider 50 MPI processes an adequate scale for our experiments. Given that the
possible sources of overhead in our approach are constant rather than dependent
on the number of nodes, nothing suggests the possibility of scalability problems
in larger experiments.

Some of the benchmarks/applications contain intra-node parallelism. In those,
the number of threads per process is 48, whereas, for the rest, \highlightForReview{it} is 1.

The runs performed for our experiments took about 10 minutes. \highlightForReview{This} means that the
whole process, from the first run until successful completion, including the
restart, took about 10 minutes. Regarding the checkpoint frequency, we forced
one checkpoint per minute, resulting in a total of ten checkpoints per run. The
frequency is expressed in terms of iterations, so that we checkpoint data every
10\% of the iterations of the program. This is a high checkpoint frequency, \highlightForReview{which was} selected on purpose to stress the checkpointing mechanisms and ease the
performance comparison between the different evaluated approaches. Coarser
checkpoint frequencies should result in even lower overheads.

\highlightForReview{Regarding} the faults, all of \highlightForReview{them} were deterministically injected
at 90\% of the application progress. The faults introduced are exceptions that
cause process abortion, and the degree of progress \highlightForReview{was} arbitrarily chosen.
The reason for an evaluation in the presence of faults is the possibility of
measuring not only the overhead introduced in checkpointing but also in the
restart process.

\subsection{Environment and applications}
In this subsection, we detail the environment \highlightForReview{where} the experiments \highlightForReview{were} run, as well
as the applications and benchmarks that we used to evaluate our approach.

The experiments were carried out on \highlightForReview{a} machine with the configuration given in
Table~\ref{tab:mn}. More details about the hardware can be found
in~\cite{mn_arch}.

\begin{table}[]
\centering
\caption{Machine architecture}
\label{tab:mn}
\begin{tabular}{@{}ll@{}}
\toprule
\textbf{Component} & \textbf{Details}         \\ \midrule
Nodes         & 3456 \\
CPU           & 2x Intel Xeon Platinum 8160 2.1 GHz  \\
Network       & 100 GB Intel Omni-Path Full-Fat Tree \\
              & \& 10G bit Ethernet \\
Memory        & 3240x 96 GB/node \& 216x 384 GB/node (Total: 384.75 TB) \\
Local storage & 240 GB Intel s3520 SSD \\
File system   & ~14 PB GPFS disk storage \\
OS            & Linux-SuSe Enterprise Server 12 SP2 \\ \bottomrule
\end{tabular}
\vspace*{-1mm}
\end{table}

The software used for our experiments, along with their versions, can be seen
in Table~\ref{tab:software}.

\begin{table}[]
\centering
\caption{Software and versions used to perform the experiments}
\label{tab:software}
\begin{tabular}{@{}ll@{}}
\toprule
\textbf{Software}                                     & \textbf{Version}                            \\ \midrule
Transparent Checkpoint Library                        & 1.0                    \\
Mercurium source-to-source compiler                   & 2.3.0                                       \\
gcc and gfortran                                      & 7.2.0                                       \\
icc and ifort                                         & 17.0.4                                      \\
Intel MPI                                             & 2017.3.196                                  \\
SCR                                                   & 1.2.2                                       \\
FTI                                                   & d54a9e0\footnotemark   \\ \bottomrule
\end{tabular}
\vspace*{-4mm}
\end{table}
\footnotetext{Hash of the commit used in the experiments.}

Following, we provide a brief explanation of the applications and benchmarks
used in the evaluation. The size of the applications \highlightForReview{ranges} from $\approx$ 500
to $\approx$ 15000 lines of code.  Note that there are 7 applications using FTI,
5 applications using SCR, and 2 applications using VeloC. This is because we did
not have the reference \highlightForReview{versions} (native FTI/SCR/VeloC) of all the applications
to compare against.

\begin{description}
    \item \textbf{BT-MZ}~\cite{bt-mz}: BT-MZ, extracted from the NAS Parallel
    Benchmarks, is a pseudo application that solves problems derived from CFD
    \highlightForReview{using} a block tri-diagonal solver. This implementation contains
    OpenMP+MPI.

    \item \textbf{Duct}~\cite{duct}: This pure MPI application, from \highlightForReview{the} CFD domain,
    performs a large eddy simulation of turbulent flow in a square \highlightForReview{duct}.

    \item \textbf{GERShWIN}~\cite{gershwin}: The GERShWIN application \highlightForReview{was}
    developed by INRIA under the umbrella of the DEEP-ER project. It studies
    human exposure to electromagnetic fields. To do so, it solves a system of
    Maxwell equations. The implementation, which contains OpenMP+MPI, presents
    some restrictions regarding the number of processes to run. Thus, it must
    be run with 48 nodes rather than 50.

    \item \textbf{Heat}: This pure MPI benchmark performs a \highlightForReview{2D heat} transfer
    simulation.

    \item \textbf{LULESH2.0}~\cite{LULESH2:changes}: \highlightForReview{This is a} C++ OpenMP+MPI sample
        application from Lawrence Livermore National Laboratory that models the
        propagation of a Sedov blast wave. The problem is formulated using a
        \highlightForReview{\mbox{three-dimensional}} unstructured mesh.

    \item \textbf{N-Body}: This benchmark, which simulates a dynamical system of
    particles, uses OpenMP+MPI. Its implementation \highlightForReview{constrains} the number of
    processes to run, so only 32 nodes \highlightForReview{were} used.

    \item \textbf{SPECFEM3D}: The SPECFEM3D application simulates seismic wave
    propagation using a Galerkin spectral element method. Its implementation
    relies on OpenMP+MPI, and presents some restrictions that force us to use
    only 32 nodes.

    \item \textbf{Stream}~\cite{stream}: Extracted from the HPC Challenge
    Benchmarks, the Stream benchmark measures the sustainable bandwidth and the
    corresponding computation rate for \highlightForReview{a} simple vector kernel. It is implemented
    using OpenMP+MPI.

    \item \textbf{TurboRVB}~\cite{turborvb}: This pure MPI application \highlightForReview{was} also
    developed under the umbrella of the DEEP-ER project, at SISSA. The
    goal of this application is to understand high-temperature superconductivity
    \highlightForReview{through} Quantum Monte Carlo simulations.

    \item \textbf{xPic}: \highlightForReview{This is a} C++ OpenMP+MPI HPC application deduced from
        iPic3D~\cite{MARKIDIS20101509}. It is designed for large scale
        production runs. xPic simulates space plasma in \highlightForReview{three-dimensional}
        parallel code.
\end{description}

\subsection{Evaluation and discussion}
\label{subsec:discussion}
As stated previously, our evaluation covers two different aspects. We want to demonstrate that our model is introducing no significant
overhead compared to using the native backend \highlightForReview{libraries} directly\highlightForReview{, and we want to} evaluate the programmability of our model.

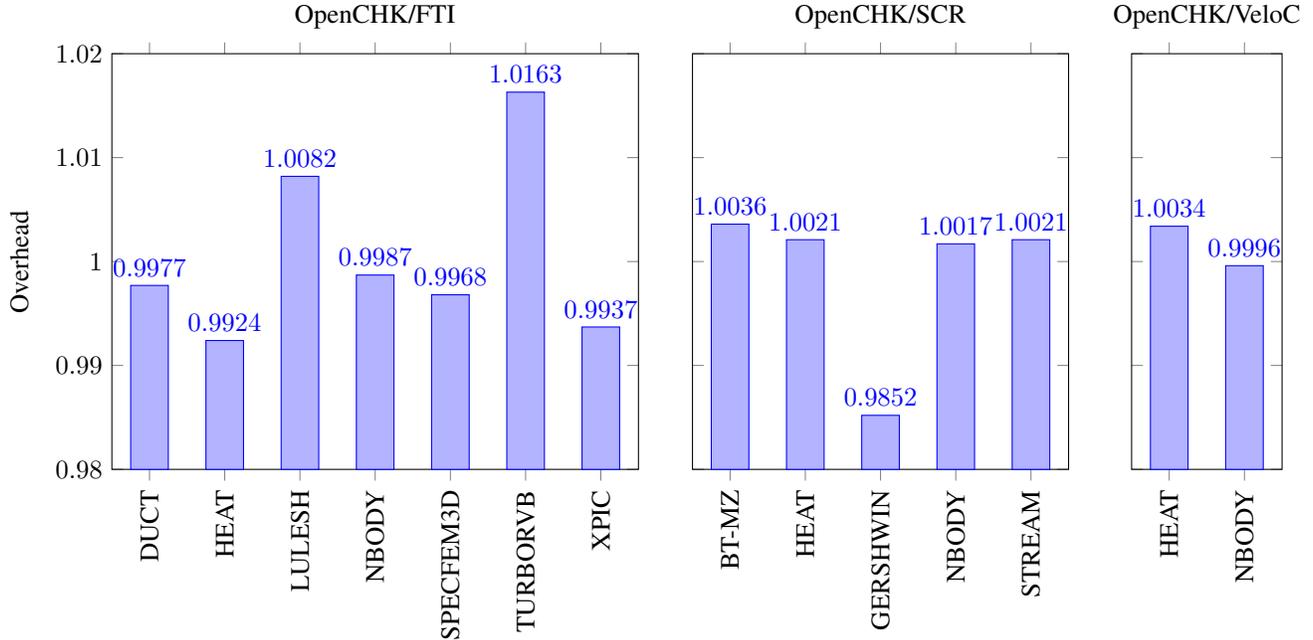
\begin{figure*}
\begin{tikzpicture}[baseline]
\begin{axis}[
    height=0.4\textwidth,
    title={OpenCHK/FTI},
    ybar,
	bar width=0.5cm,
    ylabel={Overhead},
    ymin=0.98, ymax=1.02,
	x=1.0cm,
    symbolic x coords={DUCT,HEAT,LULESH,NBODY,SPECFEM3D,TURBORVB,XPIC},
    xtick=data,
    xticklabel style={rotate=90,anchor=east,align=center},
    nodes near coords,
    nodes near coords style={/pgf/number format/.cd,precision=4},
    nodes near coords align={vertical},
    enlarge x limits={abs=0.5cm},
]
\addplot
	coordinates {(DUCT, 0.9977) (HEAT, 0.9924) (LULESH, 1.0082) (NBODY, 0.9987) (SPECFEM3D, 0.9968) (TURBORVB, 1.0163) (XPIC, 0.9937)};
\end{axis}
\end{tikzpicture}
~
\begin{tikzpicture}[baseline]
\begin{axis}[
    height=0.4\textwidth,
    title={OpenCHK/SCR},
    ybar,
	bar width=0.5cm,
    ymin=0.98, ymax=1.02,
	yticklabels={},
	x=1.0cm,
    symbolic x coords={BT-MZ,HEAT,GERSHWIN,NBODY,STREAM},
    xtick=data,
    xticklabel style={rotate=90,anchor=east,align=center},
    nodes near coords,
    nodes near coords style={/pgf/number format/.cd,precision=4},
    nodes near coords align={vertical},
    enlarge x limits={abs=0.5cm},
]
\addplot
	coordinates {(BT-MZ, 1.0036) (GERSHWIN, 0.9852) (HEAT, 1.0021) (NBODY, 1.0017) (STREAM, 1.0021)};
\end{axis}
\end{tikzpicture}
~
\begin{tikzpicture}[baseline]
\begin{axis}[
    height=0.4\textwidth,
    title={OpenCHK/VeloC},
    ybar,
	bar width=0.5cm,
    ymin=0.98, ymax=1.02,
	yticklabels={},
	x=1.0cm,
    symbolic x coords={HEAT,NBODY},
    xtick=data,
    xticklabel style={rotate=90,anchor=east,align=center},
    nodes near coords,
    nodes near coords style={/pgf/number format/.cd,precision=4},
    nodes near coords align={vertical},
    enlarge x limits={abs=0.5cm},
]
\addplot
	coordinates {(HEAT, 1.0034) (NBODY, 0.9996)};
\end{axis}
\end{tikzpicture}
\caption{Overhead introduced by \texttt{OpenCHK} \highlightForReview{compared to} using native FTI/SCR/VeloC}
\label{fig:overhead}
\end{figure*}

Regarding the first aspect of the evaluation, Figure~\ref{fig:overhead} shows
three different charts, one for each backend library. Each of the charts shows
the different applications and benchmarks executed in the \highlightForReview{x-axis}, while in the
y-axis \highlightForReview{shows} the overhead calculated as \highlightForReview{previously described}.

For the first chart starting from the left, which corresponds to FTI, it can be
seen that the differences between OpenCHK and native FTI are always $<$ 2\%. The
worst case, TurboRVB, has a difference of 1.63\%, while the rest are $<$ 1\%.
Moreover, the differences are always within the standard deviation of the runs,
which \highlightForReview{range from} $\approx$0.15\% to $\approx$2.6\%\highlightForReview{, except for TurboRVB,}
which is $\approx$4.6\%, so we conclude that negligible overhead is
introduced by OpenCHK \highlightForReview{compared to} native FTI.

The \highlightForReview{center chart}, corresponding to SCR, shows differences \highlightForReview{that are} always
$<$ 0.5\%, except for the GERShWIN application that was 1.48\%. However,
this value fits within the standard deviation (1.49\%), while the rest \highlightForReview{also remain} within their respective standard deviation values. Therefore, the overhead introduced by OpenCHK is negligible \highlightForReview{compared to} native SCR.

Finally, the right-most chart, which presents results for VeloC, exhibits
differences of $<$ 0.5\%. Furthermore, these values are within their
corresponding standard \highlightForReview{deviations}. Consequently, no
significant overhead is introduced by OpenCHK \highlightForReview{compared to} native VeloC.

Therefore, we can conclude that no significant overhead is introduced at all by the OpenCHK model when compared \highlightForReview{to} its native counterparts.

These results \highlightForReview{align with} the results published in our previous
work~\cite{maronas2017checkpoint}, \highlightForReview{which} we are extending \highlightForReview{in this work}. The evaluation of
our previous work was done in MareNostrum 3, the available machine at the
\highlightForReview{time} of that work. Now, MareNostrum has been upgraded to its fourth version,
and we have relaunched the whole application set, obtaining similar
results. It is important to highlight the differences between the total number
of cores: 64 (nodes) * 16 (CPUs per node) in MareNostrum 3, and 50 (nodes) * 48
(CPUs per node) in MareNostrum 4; \highlightForReview{the new version has more than 2x the number of processors}. However, the results
remain unchanged, demonstrating a good scalability \highlightForReview{for} our approach.

\highlightForReview{Now that we have demonstrated} that our approach introduces negligible overhead, we wish to focus
on the most important point of our approach: the programmability. We based our analysis on the SLOC metric, which stands for
\textit{source lines of code}. We measure it using SLOCCount~\cite{sloccount}.
We \highlightForReview{selected} the lines of code needed to implement each of the different
versions \highlightForReview{to} make this measurement, the results of which are shown in
Tables~\ref{tab:codelines_fti},~\ref{tab:codelines_scr},
and~\ref{tab:codelines_veloc} for FTI, SCR, and VeloC, respectively. Here it is
possible to see the lines of code required to write application-level
\highlightForReview{CR} in FTI, SCR, VeloC, and OpenCHK. The code we evaluated provides the same functionality \highlightForReview{between OpenCHK and the native versions}, but may not
be 100\% equivalent. This fact can be \highlightForReview{seen} by checking the full example of
code provided in the appendices. Moreover, native implementations include error
handling, while OpenCHK manages errors inside the TCL library.

As can be observed in Table~\ref{tab:codelines_fti}, our approach \highlightForReview{can}
drastically reduce the number of lines required to perform application-level
\highlightForReview{CR}. On average, the number of lines
required by OpenCHK \highlightForReview{was reduced to} around 30\% of the lines required by FTI to
provide the same functionality.

\begin{table}[]
\centering
\caption{Number of lines of code required to perform application-level
    CR using FTI and OpenCHK.}
\label{tab:codelines_fti}
\begin{tabular}{@{}llll@{}}
\toprule
                   & \textbf{FTI}  & \textbf{OpenCHK}  & \textbf{OpenCHK/FTI} \\ \midrule
\textbf{DUCT}      & 31            & 5                 & 0.1613             \\
\textbf{HEAT}      & 15            & 5                 & 0.3333             \\
\textbf{LULESH}    & 12            & 5                 & 0.4167             \\
\textbf{NBODY}     & 25            & 5                 & 0.2             \\
\textbf{SPECFEM3D} & 28            & 6                 & 0.2143             \\
\textbf{TURBORVB}  & 80            & 6                 & 0.075             \\
\textbf{XPIC}      & 8             & 5                 & 0.625             \\ \midrule
\textbf{AVERAGE}   & & & 0.2894           \\ \bottomrule
\end{tabular}
\end{table}

The comparison with SCR shows even better results in terms of programmability.
Table~\ref{tab:codelines_scr} shows that OpenCHK \highlightForReview{can} express
\highlightForReview{CR} in \highlightForReview{as little as} 3\% of the lines used by SCR, allowing to
express a \highlightForReview{CR} mechanism in five lines while SCR needs 165 lines
for the same purpose. The code lines needed by OpenCHK to provide the same
functionality \highlightForReview{as} SCR represents, on average, only about 6\% of those
required by SCR.

Note that these results are for new applications that do not contain output I/O
or checkpointing. For those applications that already contain code to perform
output I/O or checkpointing, our results are inflated, \highlightForReview{because} the I/O code already
exists and does not need to be added. For example, if we
assume the output code already exists, SCR involves \highlightForReview{only} 40 additional lines of code
\highlightForReview{for} the NBODY benchmark. In
general, for legacy codes and applications that have already an I/O method
implemented in the code, using SCR and/or VeloC can be more beneficial than using
a new interface because it leverages existing code; however, for freshly developed
applications or to use features of different backend checkpoint libraries, OpenCHK
does provide an easy way to access those libraries through a simple interface.

\begin{table}[]
\centering
\caption{Number of lines of code required to perform application-level
    CR using SCR and OpenCHK.}
\label{tab:codelines_scr}
\begin{tabular}{@{}llll@{}}
\toprule
                   & \textbf{SCR}  & \textbf{OpenCHK}  & \textbf{OpenCHK/SCR} \\ \midrule
\textbf{BT-MZ}     & 118           & 12                & 0.1017            \\
\textbf{GERSHWIN}  & 200           & 8                 & 0.04             \\
\textbf{HEAT}      & 78            & 5                 & 0.0641            \\
\textbf{NBODY}     & 109           & 5                 & 0.0459             \\
\textbf{STREAM}    & 165           & 5                 & 0.0303           \\ \midrule
\textbf{AVERAGE}   & & & 0.0564            \\ \bottomrule
\end{tabular}
\vspace*{-4mm}
\end{table}

As we are using VeloC in memory-based mode, the comparison results are very
similar to FTI. This is because VeloC memory-based mode is much \highlightForReview{like} FTI.
Therefore, similarly to FTI, we only need around 36\% \highlightForReview{of} the lines required by
VeloC to express \highlightForReview{CR}. If we were using VeloC in
file-based mode, as it is very similar to SCR, the comparison should be \highlightForReview{more like} SCR.

\begin{table}[]
\centering
\caption{Number of lines of code required to perform application-level CR using VeloC and OpenCHK.}
\label{tab:codelines_veloc}
\begin{tabular}{@{}llll@{}}
\toprule
                   & \textbf{VeloC}  & \textbf{OpenCHK}  & \textbf{OpenCHK/VeloC} \\ \midrule
\textbf{HEAT}      & 10            & 5                 & 0.5            \\
\textbf{NBODY}     & 23            & 5                 & 0.2174             \\
\textbf{AVERAGE}   & & & 0.3587            \\ \bottomrule
\end{tabular}
\vspace*{-4mm}
\end{table}

In general, OpenCHK usually needs only five lines to express \highlightForReview{the entire
CR} code. Two lines for initialization--- one for creating
the MPI communicator to be passed to TCL, and one for the init directive---,
another line for the load (unless there are \highlightForReview{many} variables), another line for
the store (again, unless there are \highlightForReview{many} variables), and, finally, another line
for \highlightForReview{the} shutdown directive. An additional important \highlightForReview{feature} is that OpenCHK prevents
users from modifying the natural program flow to check whether an execution is
a restart or not. Overall, in light of the results, we can conclude that
programmability is enhanced.

Finally, portability is also improved with our solution.
Users can use their OpenCHK applications with \highlightForReview{whichever} of the three backends
supported. Consequently, moving from a system with one backend (e.g., FTI) to a
system with a different backend (e.g., SCR or VeloC) requires no changes in the
source code. Otherwise, if native APIs are used, the code
related to \highlightForReview{CR must} be completely rewritten.

\section{Conclusions}
\label{sec:conclusion}
Throughout this paper, \highlightForReview{we have detailed} the extension of a directive-based approach
designed for providing application-level \highlightForReview{CR}: the OpenCHK
programming model. The model includes the new \texttt{\#pragma chk} directive 
family, composed of several directives and clauses. \highlightForReview{They allow users
to} specify data to be checkpointed persistently, along
with other details, such as checkpoint frequency, checkpoint identifier, or
checkpoint level. Additionally, the model enables users to recover data from an 
existent checkpoint, in the case of a restart after failure, continuing the 
execution from the recovered state rather than from \highlightForReview{the beginning.}

The directive-based approach presented in this paper eases the use of 
application-level \highlightForReview{CR}. The solution proposed (1) minimizes
the modifications required in the source code to perform \highlightForReview{CR}, (2) transfers the responsibility of serializing and
\highlightForReview{deserializing} data required by traditional approaches from
the user side to the model side, and (3) prevents users from modifying the
natural program flow to check whether data can be recovered from a checkpoint. 
Our solution incorporates state-of-the-art \highlightForReview{CR} libraries
(FTI, SCR, and VeloC) \highlightForReview{to} maximize resiliency and performance,
benefiting from their advanced redundancy schemes and their highly optimized 
I/O operations. \highlightForReview{Additionally}, the OpenCHK model enables users to \highlightForReview{employ} any of
the backend libraries without modifying a single line of source code, thereby 
enhancing the portability of applications. The OpenCHK model can be combined 
with other programming models, such as OpenMP, MPI, and other similar programming
models like OmpSs, as well as combinations of them.

Furthermore, the OpenCHK model not only supports the basic
functionality but also advanced functionalities: CP-dedicated threads to reduce
checkpointing overhead in some architectures, differential checkpoints to store
only the blocks of data that have been modified, saving time and space, and
support for HDF5 to allow merging \highlightForReview{CR} with data analytics.
Moreover, given its nature, OpenCHK is easily extensible so that new features
implemented in any of the backends can be added to the model. Our contribution
consists not only of the model, but also an implementation. Our
implementation provides robust features to help users increase their
productivity, \highlightForReview{such as} self-iterative data expressions, \highlightForReview{which} are useful when dealing with
arrays to avoid tedious and error-prone tasks.

Our evaluation, consisting \highlightForReview{of} several benchmarks and production-level scientific
applications, showed (1) no significant overhead compared to using the native 
APIs of state-of-the-art solutions such as FTI, SCR, and VeloC, and (2) a significant
reduction of the required number of source code lines to perform 
application-level \highlightForReview{CR}. On average, OpenCHK needs only 29\%,
6\%, and 36\% of the code lines required by FTI, SCR, and VeloC,
respectively, to perform the same functionalities. Finally, we enhanced 
portability, enabling users to choose among FTI, SCR, or VeloC at runtime,
with no changes in the source code.

\section{Future work}
\label{sec:future}

As future work, we would like to integrate incremental checkpoints \highlightForReview{into OpenCHK}. This
is a technique where \highlightForReview{a} checkpoint is not fully written \highlightForReview{at} one time, but
incrementally built in several separated write operations. An example of this
is an N-body simulation dealing with particle positions, velocities, and forces.
Each one of these is calculated at \highlightForReview{a} different time, starting \highlightForReview{with} the forces, then
the velocities, and finally the positions. When the forces have been updated,
they can be written in the checkpoint, possibly \highlightForReview{while the}
velocities are being calculated. Then, when the velocities have been updated,
they can be written in the checkpoint, and the same finally with the positions.
Overall, all the variables are checkpointed, but the write operations 
are separated in time, to decrease storage congestion and maximize 
parallelization.

Another idea is decoupling the actual operation (load/store) and the data
registration. Currently, the model does \highlightForReview{these together because} the data is
registered in the load/store clauses. However, it may become a problem when
dealing with C++ classes due to the visibility of some members in different
contexts. Therefore, allowing registration and actual load/store separately
would help in some specific cases.

Finally, we plan to add GPU checkpointing to the model \highlightForReview{to} accelerate
fault-tolerance tasks and \highlightForReview{better exploit} the resources of heterogeneous
systems.

\section*{Acknowledgment}

This work is supported by the Spanish Ministerio de Ciencia, Innovaci\'{o}n y
Universidades (TIN2015-65316-P) and by the Generalitat de Catalunya
(2014-SGR-1051). This project received funding from the European
Union's Seventh Framework Programme (FP7/2007-2013) and the Horizon 2020
(H2020) funding framework under grant agreement no.  H2020-FETHPC-754304
(DEEP-EST).  The present publication reflects only the authors' views. The
European Commission is not liable for any use that might be made of the
information contained herein. We would also like to
thank the reviewers for their feedback and contributions to this work.

\bibliographystyle{IEEEtran}
\bibliography{bib/tcl}

\begin{thebibliography}{10}
\providecommand{\url}[1]{#1}
\csname url@samestyle\endcsname
\providecommand{\newblock}{\relax}
\providecommand{\bibinfo}[2]{#2}
\providecommand{\BIBentrySTDinterwordspacing}{\spaceskip=0pt\relax}
\providecommand{\BIBentryALTinterwordstretchfactor}{4}
\providecommand{\BIBentryALTinterwordspacing}{\spaceskip=\fontdimen2\font plus
\BIBentryALTinterwordstretchfactor\fontdimen3\font minus
  \fontdimen4\font\relax}
\providecommand{\BIBforeignlanguage}[2]{{%
\expandafter\ifx\csname l@#1\endcsname\relax
\typeout{** WARNING: IEEEtran.bst: No hyphenation pattern has been}%
\typeout{** loaded for the language `#1'. Using the pattern for}%
\typeout{** the default language instead.}%
\else
\language=\csname l@#1\endcsname
\fi
#2}}
\providecommand{\BIBdecl}{\relax}
\BIBdecl

\bibitem{maronas2017checkpoint}
M.~Maro{\~n}as, S.~Mateo, V.~Beltran, and E.~Ayguad{\'e}, ``A directive-based
  approach to perform persistent checkpoint/restart,'' in \emph{2017
  International Conference on High Performance Computing Simulation (HPCS)},
  July 2017, pp. 442--451.

\bibitem{reed2004}
D.~Reed, ``High-end computing: The challenge of scale,'' Director’s
  Colloquium, Los Alamos National Laboratory, May 2004.

\bibitem{cappello2014toward}
F.~Cappello, A.~Geist, W.~Gropp, S.~Kale, B.~Kramer, and M.~Snir, ``Toward
  exascale resilience: 2014 update,'' \emph{Supercomputing frontiers and
  innovations}, vol.~1, no.~1, pp. 5--28, 2014.

\bibitem{leo}
L.~Bautista-Gomez and F.~Cappello, ``Detecting and correcting data corruption
  in stencil applications through multivariate interpolation,'' in
  \emph{International Conference on Cluster Computing (CLUSTER)}, IEEE.\hskip
  1em plus 0.5em minus 0.4em\relax Chicago, IL: IEEE, Sep. 2015, pp. 595--602.

\bibitem{du2012algorithm}
P.~Du, A.~Bouteiller, G.~Bosilca, T.~Herault, and J.~Dongarra,
  ``Algorithm-based fault tolerance for dense matrix factorizations,''
  \emph{ACM SIGPLAN Notices}, vol.~47, no.~8, pp. 225--234, 2012.

\bibitem{blcr}
P.~H. Hargrove and J.~C. Duell, ``Berkeley lab checkpoint\slash restart
  ({BLCR}) for {Linux} clusters,'' \emph{Journal of Physics: Conference
  Series}, vol.~46, pp. 494--499, 2006.

\bibitem{bautista2011fti}
L.~{Bautista-Gomez}, S.~{Tsuboi}, D.~{Komatitsch}, F.~{Cappello},
  N.~{Maruyama}, and S.~{Matsuoka}, ``Fti: High performance fault tolerance
  interface for hybrid systems,'' in \emph{SC '11: Proceedings of 2011
  International Conference for High Performance Computing, Networking, Storage
  and Analysis}, Nov 2011, pp. 1--12.

\bibitem{scr}
A.~{Moody}, G.~{Bronevetsky}, K.~{Mohror}, and B.~R. d.~{Supinski}, ``Design,
  modeling, and evaluation of a scalable multi-level checkpointing system,'' in
  \emph{SC '10: Proceedings of the 2010 ACM/IEEE International Conference for
  High Performance Computing, Networking, Storage and Analysis}, Nov 2010, pp.
  1--11.

\bibitem{veloc}
\BIBentryALTinterwordspacing
{Argonne National Laboratory and Lawrence Livermore National Laboratory},
  ``{VELOC},'' accessed: 2019-03-24. [Online]. Available:
  \url{https://veloc.readthedocs.io/en/latest/toc.html}
\BIBentrySTDinterwordspacing

\bibitem{openchk}
\BIBentryALTinterwordspacing
{Barcelona Supercomputing Center}, ``{OpenCHK},'' accessed: 2019-03-24.
  [Online]. Available: \url{https://github.com/bsc-pm/OpenCHK-model}
\BIBentrySTDinterwordspacing

\bibitem{openmp}
\BIBentryALTinterwordspacing
{OpenMP Architecture Review Board}, ``{OpenMP Application Programming
  Interface},'' November 2018, accessed: 2019-03-24. [Online]. Available:
  \url{https://www.openmp.org/wp-content/uploads/OpenMP-API-Specification-5.0.pdf}
\BIBentrySTDinterwordspacing

\bibitem{stellner1996cocheck}
G.~Stellner, ``Cocheck: Checkpointing and process migration for mpi,'' in
  \emph{Proceedings of International Conference on Parallel Processing}.\hskip
  1em plus 0.5em minus 0.4em\relax IEEE, 1996, pp. 526--531.

\bibitem{bronevetsky2003automated}
G.~Bronevetsky, D.~Marques, K.~Pingali, and P.~Stodghill, ``Automated
  application-level checkpointing of mpi programs,'' in \emph{ACM Sigplan
  Notices}, vol.~38, no.~10.\hskip 1em plus 0.5em minus 0.4em\relax ACM, 2003,
  pp. 84--94.

\bibitem{plank1998diskless}
J.~S. Plank, K.~Li, and M.~A. Puening, ``Diskless checkpointing,'' \emph{IEEE
  Transactions on Parallel and Distributed Systems}, vol.~9, no.~10, pp.
  972--986, 1998.

\bibitem{zheng2004ftc}
G.~Zheng, L.~Shi, and L.~V. Kal{\'e}, ``Ftc-charm++: an in-memory
  checkpoint-based fault tolerant runtime for charm++ and mpi,'' in \emph{2004
  ieee international conference on cluster computing (ieee cat. no.
  04EX935)}.\hskip 1em plus 0.5em minus 0.4em\relax IEEE, 2004, pp. 93--103.

\bibitem{duell2002requirements}
\BIBentryALTinterwordspacing
J.~Duell, P.~H. Hargrove, and E.~S. Roman. (2002, Feb.) Requirements for linux
  checkpoint/restart. [Online]. Available:
  \url{http://escholarship.org/uc/item/0rh4s63j}
\BIBentrySTDinterwordspacing

\bibitem{roman2002survey}
E.~Roman, ``A survey of checkpoint/restart implementations,'' Lawrence Berkeley
  National Laboratory, Tech, Tech. Rep., 2002.

\bibitem{duell2005design}
\BIBentryALTinterwordspacing
J.~Duell. (2005, Apr.) The design and implementation of berkely lab's linux
  checkpoint/restart. [Online]. Available:
  \url{https://escholarship.org/uc/item/40v987j0}
\BIBentrySTDinterwordspacing

\bibitem{sankaran2005lam}
S.~Sankaran, J.~M. Squyres, B.~Barrett, V.~Sahay, A.~Lumsdaine, J.~Duell,
  P.~Hargrove, and E.~Roman, ``The lam/mpi checkpoint/restart framework:
  System-initiated checkpointing,'' \emph{International Journal of High
  Performance Computing Applications}, vol.~19, no.~4, pp. 479--493, 2005.

\bibitem{bronevetsky2004application}
G.~Bronevetsky, D.~Marques, K.~Pingali, P.~Szwed, and M.~Schulz,
  ``Application-level checkpointing for shared memory programs,'' \emph{ACM
  SIGARCH Computer Architecture News}, vol.~32, no.~5, pp. 235--247, 2004.

\bibitem{young1974first}
J.~W. Young, ``A first order approximation to the optimum checkpoint
  interval,'' \emph{Communications of the ACM}, vol.~17, no.~9, pp. 530--531,
  1974.

\bibitem{duda1983effects}
A.~Duda, ``The effects of checkpointing on program execution time,''
  \emph{Information Processing Letters}, vol.~16, no.~5, pp. 221--229, 1983.

\bibitem{plank2001processor}
J.~S. Plank and M.~G. Thomason, ``Processor allocation and checkpoint interval
  selection in cluster computing systems,'' \emph{Journal of Parallel and
  distributed Computing}, vol.~61, no.~11, pp. 1570--1590, 2001.

\bibitem{daly2006higher}
J.~T. Daly, ``A higher order estimate of the optimum checkpoint interval for
  restart dumps,'' \emph{Future Generation Computer Systems}, vol.~22, no.~3,
  pp. 303--312, 2006.

\bibitem{roadrunnertechreport2007}
G.~Grider, J.~Loncaric, and D.~Limpert, ``Roadrunner system management
  report,'' Los Alamos National Laboratory, Tech. Rep. LA-UR-07-7405, 2007.

\bibitem{schroeder2007}
\BIBentryALTinterwordspacing
B.~Schroeder and G.~A. Gibson, ``Understanding failures in petascale
  computers,'' \emph{Journal of Physics: Conference Series}, vol.~78, no.~1, p.
  012022, 2007. [Online]. Available:
  \url{http://stacks.iop.org/1742-6596/78/i=1/a=012022}
\BIBentrySTDinterwordspacing

\bibitem{oldfield2007modeling}
R.~A. Oldfield, S.~Arunagiri, P.~J. Teller, S.~Seelam, M.~R. Varela, R.~Riesen,
  and P.~C. Roth, ``Modeling the impact of checkpoints on next-generation
  systems,'' in \emph{24th IEEE Conference on Mass Storage Systems and
  Technologies (MSST 2007)}.\hskip 1em plus 0.5em minus 0.4em\relax San Diego,
  CA: IEEE, Sept 2007, pp. 30--46.

\bibitem{schroeder2010large}
B.~Schroeder and G.~A. Gibson, ``A large-scale study of failures in
  high-performance computing systems,'' \emph{Dependable and Secure Computing,
  IEEE Transactions on}, vol.~7, no.~4, pp. 337--350, 2010.

\bibitem{gelenbe1976model}
\BIBentryALTinterwordspacing
E.~Gelenbe, ``A model of roll-back recovery with multiple checkpoints,'' in
  \emph{Proceedings of the 2Nd International Conference on Software
  Engineering}, ser. ICSE '76.\hskip 1em plus 0.5em minus 0.4em\relax Los
  Alamitos, CA, USA: IEEE Computer Society Press, 1976, pp. 251--255. [Online].
  Available: \url{http://dl.acm.org/citation.cfm?id=800253.807684}
\BIBentrySTDinterwordspacing

\bibitem{vaidya1994case}
N.~H. Vaidya, ``A case for multi-level distributed recovery schemes,'' -, Tech.
  Rep., 1994.

\bibitem{pbkl:95:lib}
J.~S. Plank, M.~Beck, G.~Kingsley, and K.~Li, ``{\bf Libckpt}: Transparent
  checkpointing under {Unix},'' in \emph{Usenix Winter Technical Conference},
  January 1995, pp. 213--223.

\bibitem{BautistaKellerDcp2019}
K.~{Keller} and L.~{Bautista-Gomez}, ``Application-level differential
  checkpointing for hpc applications with dynamic datasets,'' in \emph{CCGrid
  '19}, May 2019.

\bibitem{hdf5}
\BIBentryALTinterwordspacing
{The HDF Group}, ``{HDF5 User's Guide},'' September 2019. [Online]. Available:
  \url{https://portal.hdfgroup.org/display/HDF5/HDF5+User%27s+Guide}
\BIBentrySTDinterwordspacing

\bibitem{mercurium}
\BIBentryALTinterwordspacing
{Barcelona Supercomputing Center}, ``{Mercurium Compiler},'' accessed:
  2019-03-24. [Online]. Available: \url{https://github.com/bsc-pm/mcxx}
\BIBentrySTDinterwordspacing

\bibitem{tcl}
\BIBentryALTinterwordspacing
{Barcelona Supercomputing Center }, ``{Transparent Checkpoint Library (TCL)},''
  accessed: 2019-03-24. [Online]. Available:
  \url{https://github.com/bsc-pm/TCL}
\BIBentrySTDinterwordspacing

\bibitem{lizard}
\BIBentryALTinterwordspacing
{Yin, Terry}, ``{Lizard},'' accessed: 2019-03-24. [Online]. Available:
  \url{https://github.com/terryyin/lizard}
\BIBentrySTDinterwordspacing

\bibitem{sloccount}
\BIBentryALTinterwordspacing
{Wheeler, David A.}, ``{SLOCCount},'' accessed: 2019-03-24. [Online].
  Available: \url{https://dwheeler.com/sloccount/}
\BIBentrySTDinterwordspacing

\bibitem{mn_arch}
\BIBentryALTinterwordspacing
{Barcelona Supercomputing Center}, ``{MareNostrum IV User's Guide},''
  Barcelona, Spain, March 2019. [Online]. Available:
  \url{https://www.bsc.es/support/MareNostrum4-ug.pdf}
\BIBentrySTDinterwordspacing

\bibitem{bt-mz}
\BIBentryALTinterwordspacing
{NASA}, ``{NAS Parallel Benchmarks},'' accessed: 2019-03-24. [Online].
  Available: \url{http://www.nas.nasa.gov/publications/npb.html}
\BIBentrySTDinterwordspacing

\bibitem{duct}
\BIBentryALTinterwordspacing
{Chernousov, Andrei}, ``{Free CFD Source Codes},'' accessed: 2019-03-24.
  [Online]. Available:
  \url{http://www.geocities.ws/MotorCity/Pit/9939/freecfd.htm}
\BIBentrySTDinterwordspacing

\bibitem{gershwin}
\BIBentryALTinterwordspacing
{INRIA}, ``{Human Exposure to Electromagnetic Fields},'' accessed: 2019-03-24.
  [Online]. Available:
  \url{https://www.deep-projects.eu/applications/project-applications/human-exposure-to-electromagnetic-fields.html}
\BIBentrySTDinterwordspacing

\bibitem{LULESH2:changes}
I.~Karlin, J.~Keasler, and R.~Neely, ``Lulesh 2.0 updates and changes,'' Tech.
  Rep. LLNL-TR-641973, August 2013.

\bibitem{stream}
\BIBentryALTinterwordspacing
{University of Tennessee}, ``{HPC Challenge Benchmark},'' accessed: 2019-03-24.
  [Online]. Available: \url{http://icl.cs.utk.edu/hpcc/}
\BIBentrySTDinterwordspacing

\bibitem{turborvb}
\BIBentryALTinterwordspacing
{CINECA}, ``{High Temperature Superconductivity},'' accessed: 2019-03-24.
  [Online]. Available:
  \url{http://www.deep-er.eu/applications/project-applications/high-temperature-superconductivity.html}
\BIBentrySTDinterwordspacing

\bibitem{MARKIDIS20101509}
\BIBentryALTinterwordspacing
S.~Markidis, G.~Lapenta, and Rizwan-uddin, ``Multi-scale simulations of plasma
  with ipic3d,'' \emph{Mathematics and Computers in Simulation}, vol.~80,
  no.~7, pp. 1509 -- 1519, 2010, multiscale modeling of moving interfaces in
  materials. [Online]. Available:
  \url{http://www.sciencedirect.com/science/article/pii/S0378475409002444}
\BIBentrySTDinterwordspacing

\end{thebibliography}

\begin{appendices}
\onecolumn
\newcommand{\Hilight}{\makebox[0pt][l]{\color{cyan}\rule[-4pt]{0.85\linewidth}{14pt}}}
\section{Full example of N-body simulation kernel using OpenCHK}
\label{sec:appendixA}
\begin{figure}[h]
\begin{lstlisting}[escapechar=$]
void solve_nbody(particles_block_t * local, 
                 particles_block_t * tmp, 
                 force_block_t * forces,
                 const int n_blocks, 
                 const int timesteps, 
                 const float time_interval)
{
   int rank, rank_size;
   MPI_Comm_rank(MPI_COMM_WORLD, &rank);
   MPI_Comm_size(MPI_COMM_WORLD, &rank_size);

   int t = 0;

   // Load local and t vars, if any.
   $\Hilight$#pragma chk load ([n_blocks] local, t)

   for (; t < timesteps; t++) {
       #pragma oss task inout([n_blocks] local)
       {
           // Store local and t vars, using t as id.
           // Each checkpoint done must be at level 4.
           // Checkpoint every 10 iterations.
           $\Hilight$#pragma chk store ([n_blocks] local, t) id (t) level (4) \
           $\Hilight$            kind (CHK_FULL) if (t%(timesteps/10)==0 && t != 0)
       }

       particles_block_t * remote = local;
       for(int i=0; i < rank_size; i++){
           #pragma oss task in([n_blocks] local,[n_blocks] remote) \
                            inout([n_blocks] forces)
           calculate_forces(forces, local, remote, n_blocks);

           #pragma oss task in([n_blocks] remote) \
                            out([n_blocks] tmp)
           exchange_particles(remote, tmp, n_blocks, rank, rank_size, i, t);

           remote=tmp;
       }

       #pragma oss task inout([n_blocks] local) inout([n_blocks] forces)
       update_particles(n_blocks, local, forces, time_interval);
   }
   #pragma oss taskwait
}
\end{lstlisting}
\caption{Full example of N-body simulation kernel using OpenCHK.}
\end{figure}
\pagebreak

\section{Full example of N-body simulation kernel using FTI}
\label{sec:appendixB}
\begin{figure}[h]
\begin{lstlisting}[escapechar=$]
$\Hilight$#include <fti.h>

void solve_nbody(particles_block_t * local, 
                 particles_block_t * tmp, 
                 force_block_t * forces,
                 const int n_blocks, 
                 const int timesteps, 
                 const float time_interval)
{
    int rank, rank_size;
    MPI_Comm_rank(MPI_COMM_WORLD, &rank);
    MPI_Comm_size(MPI_COMM_WORLD, &rank_size);

    int t = 0;
    // Create a new FTI data type
    $\Hilight$FTIT_type ckptInfo;
    // Initialize the new FTI data type
    $\Hilight$FTI_InitType(&ckptInfo, n_blocks*sizeof(particles_block_t));
    $\Hilight$FTI_Protect(0, &t, sizeof(int), FTI_INTG);
    $\Hilight$FTI_Protect(1, local, 1, ckptInfo);

    $\Hilight$if( FTI_Status() ) {
    $\Hilight$    FTI_Recover();
    $\Hilight$}
    for (; t < timesteps; t++) {
        #pragma oss task inout([n_blocks] local)
        {
            $\Hilight$if(t%(timesteps/10)==0 && t != 0 && restarted_t != 0 && t!=restarted_t) {
            $\Hilight$    int res = FTI_Checkpoint( t, 4 );
            $\Hilight$    if( res != FTI_DONE ) {
            $\Hilight$        printf( "FTI internal error." );
            $\Hilight$        MPI_Abort( MPI_COMM_WORLD, -1 );
            $\Hilight$    }
            $\Hilight$}
        }

       particles_block_t * remote = local;
       for(int i=0; i < rank_size; i++){
           #pragma oss task in([n_blocks] local,[n_blocks] remote) \
                            inout([n_blocks] forces)
           calculate_forces(forces, local, remote, n_blocks);

           #pragma oss task in([n_blocks] remote) \
                            out([n_blocks] tmp)
           exchange_particles(remote, tmp, n_blocks, rank, rank_size, i, t);

           remote=tmp;
       }

       #pragma oss task inout([n_blocks] local) inout([n_blocks] forces)
       update_particles(n_blocks, local, forces, time_interval);
    }

    #pragma oss taskwait
}
\end{lstlisting}
\caption{Full example of N-body simulation kernel using FTI.}
\end{figure}
\pagebreak

\section{Full example of N-body simulation kernel using VeloC}
\label{sec:appendixC}
\begin{figure}[h]
\begin{lstlisting}[escapechar=$]
$\Hilight$#include <veloc.h>

void solve_nbody(particles_block_t * local, 
                 particles_block_t * tmp, 
                 force_block_t * forces,
                 const int n_blocks, 
                 const int timesteps, 
                 const float time_interval)
{
    int rank, rank_size;
    MPI_Comm_rank(MPI_COMM_WORLD, &rank);
    MPI_Comm_size(MPI_COMM_WORLD, &rank_size);

    int t = 0;
    $\Hilight$VELOC_Mem_protect(0, &t, 1, sizeof(int));
    $\Hilight$VELOC_Mem_protect(1, local, n_blocks, sizeof(particles_block_t));


    $\Hilight$int restarted_version = VELOC_Restart_test("nbody", t);
    $\Hilight$if(restarted_version != VELOC_FAILURE) {
    $\Hilight$    assert(VELOC_Restart("nbody", restarted_version) == VELOC_SUCCESS);
    $\Hilight$}
    for (; t < timesteps; t++) {
        #pragma oss task inout([n_blocks] local)
        {
            $\Hilight$if(t%(timesteps/10)==0 && t != 0 && restarted_t != 0 && t!=restarted_t) {
            $\Hilight$    int res = VELOC_Checkpoint("nbody", t);
            $\Hilight$    if( res != VELOC_SUCCESS ) {
            $\Hilight$        printf( "VELOC internal error." );
            $\Hilight$        MPI_Abort( MPI_COMM_WORLD, -1 );
            $\Hilight$    }
            $\Hilight$}
        }

       particles_block_t * remote = local;
       for(int i=0; i < rank_size; i++){
           #pragma oss task in([n_blocks] local,[n_blocks] remote) \
                            inout([n_blocks] forces)
           calculate_forces(forces, local, remote, n_blocks);

           #pragma oss task in([n_blocks] remote) \
                            out([n_blocks] tmp)
           exchange_particles(remote, tmp, n_blocks, rank, rank_size, i, t);

           remote=tmp;
       }

       #pragma oss task inout([n_blocks] local) inout([n_blocks] forces)
       update_particles(n_blocks, local, forces, time_interval);
    }

    #pragma oss taskwait
}
\end{lstlisting}
\caption{Full example of N-body simulation kernel using VeloC.}
\end{figure}
\pagebreak

\section{Full example of N-body simulation kernel using SCR}
\label{sec:appendixD}
\begin{figure}[h]
\begin{lstlisting}[escapechar=$]
$\Hilight$#include <scr.h>

void solve_nbody_cp(const int n_blocks, 
                    const int rank, 
                    particles_block_t const* __restrict__ local, 
                    const int timestep);
int  solve_nbody_rt(const int n_blocks, 
                    const int rank, 
                    particles_block_t* __restrict__ local, 
                    int *timestep);

void solve_nbody(particles_block_t * local, 
                 particles_block_t * tmp, 
                 force_block_t * forces,
                 const int n_blocks, 
                 const int timesteps, 
                 const float time_interval)
{
    int rank, rank_size;
    MPI_Comm_rank(MPI_COMM_WORLD, &rank);
    MPI_Comm_size(MPI_COMM_WORLD, &rank_size);

    int t = 0;
    $\Hilight$solve_nbody_rt(n_blocks, rank, local, &t);
    for (; t < timesteps; t++) {
        #pragma oss task inout([n_blocks] local)
        {
            if(t%(timesteps/10)==0 && t != 0 && restarted_t != 0 && t!=restarted_t) {
                $\Hilight$solve_nbody_cp(n_blocks, rank, local, t);
            }
        }

       particles_block_t * remote = local;
       for(int i=0; i < rank_size; i++){
           #pragma oss task in([n_blocks] local,[n_blocks] remote) \
                            inout([n_blocks] forces)
           calculate_forces(forces, local, remote, n_blocks);

           #pragma oss task in([n_blocks] remote) \
                            out([n_blocks] tmp)
           exchange_particles(remote, tmp, n_blocks, rank, rank_size, i, t);

           remote=tmp;
       }

       #pragma oss task inout([n_blocks] local) inout([n_blocks] forces)
       update_particles(n_blocks, local, forces, time_interval);
    }

    #pragma oss taskwait
}
\end{lstlisting}
\caption{First part of full example of N-body simulation kernel using SCR.}
\end{figure}

\begin{figure}[h]
\begin{lstlisting}[escapechar=$]
$\Hilight$void solve_nbody_cp(const int n_blocks, 
$\Hilight$                    const int rank,
$\Hilight$                    particles_block_t const* __restrict__ local, 
$\Hilight$                    const int timestep) {
$\Hilight$    int status;
$\Hilight$    int saved_data = 0;
$\Hilight$    int saved_data_2 = 0;
$\Hilight$    int i=0;
$\Hilight$    int res;
$\Hilight$
$\Hilight$    char name[256];
$\Hilight$    char path[SCR_MAX_FILENAME];
$\Hilight$
$\Hilight$    int perform_checkpoint;
$\Hilight$    SCR_Need_checkpoint(&perform_checkpoint);
$\Hilight$    if (perform_checkpoint == 1) {
$\Hilight$        res = SCR_Start_checkpoint();
$\Hilight$        if(res != SCR_SUCCESS)
$\Hilight$            assert(0 && "SCR failed starting a checkpoint.");
$\Hilight$
$\Hilight$        const char * scr_prefix = getenv("SCR_PREFIX");
$\Hilight$        sprintf(name, "%s/solve_nbody-nb%d_bs%d_r%d.ckpt", scr_prefix, n_blocks, BLOCK_SIZE, rank);
$\Hilight$
$\Hilight$        // Get backup file path
$\Hilight$        if(SCR_Route_file(name, path)==SCR_SUCCESS){
$\Hilight$            cint fd = open (path, 
$\Hilight$                            O_WRONLY | O_CREAT | O_TRUNC, 
$\Hilight$                            S_IRUSR | S_IRGRP | S_IROTH | S_IWUSR | S_IWGRP | S_IWOTH);
$\Hilight$
$\Hilight$            // Open, write and close file
$\Hilight$            assert(fd >= 0);
$\Hilight$            saved_data = write(fd, &timestep, sizeof(int));
$\Hilight$            saved_data_2 = write(fd, local, sizeof(particles_block_t)*n_blocks);
$\Hilight$            assert(close(fd)==0);
$\Hilight$        }
$\Hilight$        int is_valid = (saved_data+saved_data_2) == (sizeof(particles_block_t)*n_blocks + sizeof(int));
$\Hilight$        SCR_Complete_checkpoint(is_valid);
$\Hilight$    }
$\Hilight$}
\end{lstlisting}
\caption{Second part of full example of N-body simulation kernel using SCR.}
\end{figure}
\pagebreak

\begin{figure}[h]
\begin{lstlisting}[escapechar=$]
$\Hilight$int solve_nbody_rt(const int n_blocks, 
$\Hilight$                   const int rank, 
$\Hilight$                   particles_block_t* __restrict__ local, 
$\Hilight$                   int *current_timestep) {
$\Hilight$    int status = 0, found_cp = 0, temp_tstep, num_read, num_read_particles, size;
$\Hilight$    MPI_Comm_size(MPI_COMM_WORLD, &size);
$\Hilight$
$\Hilight$    char name[256], path[SCR_MAX_FILENAME], cp_name[SCR_MAX_FILENAME];
$\Hilight$    sprintf(name, "solve_nbody-nb%d_bs%d_r%d.ckpt", n_blocks, BLOCK_SIZE, rank);
$\Hilight$
$\Hilight$    int res = SCR_Have_restart(&found_cp, path);
$\Hilight$    if(res != SCR_SUCCESS)
$\Hilight$        assert(0 && "SCR failed when checking for available restarts.");
$\Hilight$
$\Hilight$    if(!found_cp)
$\Hilight$        return -1;
$\Hilight$    found_cp = 0;
$\Hilight$
$\Hilight$    if(SCR_Start_restart(cp_name) != SCR_SUCCESS)
$\Hilight$        assert(0 && "SCR failed starting a restart.");
$\Hilight$
$\Hilight$    if (SCR_Route_file(name, path) == SCR_SUCCESS) {
$\Hilight$        cint fd = open (path, 
$\Hilight$                        O_RDWR | O_CREAT, 
$\Hilight$                        S_IRUSR | S_IRGRP | S_IROTH | S_IWUSR | S_IWGRP | S_IWOTH);
$\Hilight$        assert(fd >= 0);
$\Hilight$
$\Hilight$        num_read = read(fd, &temp_tstep, sizeof(int));
$\Hilight$        void *tmp = (void *) mmap(NULL, 
$\Hilight$                                  n_blocks*sizeof(particles_block_t)+sizeof(int), 
$\Hilight$                                  PROT_WRITE|PROT_READ, MAP_SHARED, fd, 0);
$\Hilight$        assert(close(fd)==0);
$\Hilight$
$\Hilight$        if(num_read == sizeof(int) && tmp != MAP_FAILED) {
$\Hilight$            found_cp = 1;
$\Hilight$            char * aux = (char *) tmp + sizeof(int);
$\Hilight$            memcpy( local, aux, n_blocks*sizeof(particles_block_t ));
$\Hilight$            munmap( tmp, n_blocks*sizeof(particles_block_t) );
$\Hilight$        } else {
$\Hilight$            status = -1;
$\Hilight$        }
$\Hilight$    }
$\Hilight$    
$\Hilight$    // Follows in next figure.
\end{lstlisting}
\caption{Third part of full example of N-body simulation kernel using SCR.}
\end{figure}

\begin{figure}
\begin{lstlisting}[escapechar=$]
$\Hilight$
$\Hilight$    if(SCR_Complete_restart(found_cp) != SCR_SUCCESS)
$\Hilight$        status = -5;
$\Hilight$
$\Hilight$    /* determine whether all tasks successfully read their checkpoint file */
$\Hilight$    int all_found_checkpoint = 0;
$\Hilight$    MPI_Allreduce(&found_cp, &all_found_checkpoint, 1, MPI_INT, MPI_LAND,
$\Hilight$            MPI_COMM_WORLD);
$\Hilight$    if (!all_found_checkpoint)
$\Hilight$        status = -2;
$\Hilight$
$\Hilight$    /* check that everyone is at the same timestep */
$\Hilight$    int timestep_and, timestep_or;
$\Hilight$    MPI_Allreduce(current_timestep, &timestep_and, 1, MPI_INT, MPI_BAND, MPI_COMM_WORLD);
$\Hilight$    MPI_Allreduce(current_timestep, &timestep_or, 1, MPI_INT, MPI_BOR, MPI_COMM_WORLD);
$\Hilight$    if (timestep_and != timestep_or)
$\Hilight$        status = -3;
$\Hilight$
$\Hilight$    if(status == 0)
$\Hilight$        *current_timestep = temp_tstep;
$\Hilight$
$\Hilight$    return status;
$\Hilight$}
\end{lstlisting}
\caption{Last part of full example of N-body simulation kernel using SCR.}
\end{figure}
\pagebreak

\end{appendices}

\end{document}